\pdfoutput=1
\documentclass{aa}  
\usepackage{graphicx}
\usepackage[varg]{txfonts}
\usepackage{natbib}
\bibpunct{(}{)}{;}{a}{}{,} 

\defcitealias{makelaetal14}{Paper~I}
\defcitealias{gahmetal13}{G13}
\defcitealias{gahmetal07}{G07}
\defcitealias{bessellbrett88}{BB88}

\newcommand{\twco}{{\hbox {\ensuremath{\mathrm{^{12}CO}} }}}

\newcommand{\thco}{{\hbox {\ensuremath{\mathrm{^{13}CO}} }}}

\newcommand{\kmps}{\ensuremath{\mathrm{km\,s^{-1}}}}
\newcommand{\Msun}{\ensuremath{\mathrm{M}_\odot}}

\newcommand{\mum}{\mu {\rm m}}
\def\Ks{\hbox{$K$s}}

\def\Js{\hbox{$J$s}}
\def\J{\hbox{$J$}}
\def\H{\hbox{$H$}}
\newcommand{\umag}{{$^{m}$}}
 
\def\jshks{\hbox{$J$s$\!H\!K$s}} 
\def\jhks{\hbox{$J\!H\!K$s}}              
\newcommand{\Htwo}{{\hbox {\ensuremath{\mathrm{H_2}}}}}

\begin{document}
   \title{Rosette nebula globules: Seahorse giving birth to a star
     \thanks{Based on observations done at the European Southern
       Observatory, La Silla, Chile (ESO program 088.C-0630) and
       with Apex (program O-088.F-9318).}\fnmsep
     \thanks{Table 1 is only available in electronic form at the CDS via anonymous ftp to cdsarc.u-strasbg.fr (130.79.128.5) or via http://cdsweb.u-strasbg.fr/cgi-bin/qcat?J/A+A/. The APEX CO spectra are available as a FITS file at the same address.}
}

   \author{M. M. M\"akel\"a
          \inst{1,2}
          \and
          L. K. Haikala\inst{3,4}
          \and
          G. F. Gahm\inst{5}
          }

   \institute{Department of Physics, Division of Geophysics and Astronomy,
             P.O. Box 64, FI-00014 University of Helsinki, Finland
         \and
             Dr. Karl Remeis-Sternwarte and Erlangen Center for Astroparticle Physics, Universit\"at Erlangen-N\"urnberg, D-96049 Bamberg, Germany\\
              \email{minja.maekelae@fau.de}
         \and
             Universidad de Atacama, Copayapu 485, Copiapo, Chile 
         \and
             Finnish Centre for Astronomy with ESO (FINCA),
              University of Turku, V\"ais\"al\"antie 20, 21500 Piikki\"o, Finland 
         \and
             Department of Astronomy, AlbaNova University Center, Stockholm University, Sweden
             }

    \date{}

 
  \abstract
   {The Rosette nebula is an \ion{H}{ii} region ionized mainly by the stellar cluster NGC~2244. Elephant trunks, globules, and globulettes are seen at the interface where the \ion{H}{ii} region and the surrounding molecular shell meet.}
   {We have observed a field in the northwestern part of the Rosette nebula where we study the small globules protruding from the shell. Our aim is to measure their properties and study their star-formation history in continuation of our earlier study of the features of the region.}
   %
   {We imaged the region in broadband near-infrared (NIR) \jshks\ filters and narrowband \Htwo\ 1--0 S(1), P$\beta$, and continuum filters using the SOFI camera at the ESO/NTT. The imaging was used to study the stellar population and surface brightness, create visual extinction maps, and locate star formation. Mid-infrared (MIR) \textit{Spitzer} IRAC and WISE and optical NOT images were used to further study the star formation and the structure of the globules. The NIR and MIR observations indicate an outflow, which is confirmed with CO observations made with APEX. }
   {The globules have mean number densities of $\sim4.6\times10^{4} \mathrm{cm}^{-3}$. P$\beta$ is seen in absorption in the cores of the globules where we measure visual extinctions of 11--16\umag. The shell and the globules have bright rims in the observed bands. In the \Ks\ band 20 to 40\% of the emission is due to fluorescent emission in the 2.12~$\mum$ \Htwo\ line similar to the tiny dense globulettes we studied earlier in a nearby region. We identify several stellar NIR excess candidates and four of them are also detected in the Spitzer IRAC 8.0~$\mum$ image and studied further. We find an outflow with a cavity wall bright in the 2.124~$\mum$ \Htwo\ line and at 8.0~$\mum$ in one of the globules. The outflow originates from a Class~I young stellar object (YSO) embedded deep inside the globule. An H$\alpha$ image suggests the YSO drives a possible parsec-scale outflow. Despite the morphology of the globule, the outflow does not seem to run inside the dusty fingers extending from the main globule body.}
   {}


   \maketitle

\authorrunning{M. M. M\"akel\"a}

%

\section{Introduction} \label{sect:introduction}

OB associations that consist of hot, massive O and B stars are formed in dense giant molecular clouds. The ultraviolet (UV) photons from the newly formed stars typically form an \ion{H}{ii} region that compresses the surrounding dust and molecular gas into a shell that surrounds the OB association. The OB association may trigger the formation of new OB associations \citep{blaauw64} or star formation in the same molecular cloud \citep[e.g.,][]{elmegreenlada77, smithetal10b}. Along with the massive OB stars a large number of less massive stars are formed in the associations. 
Several mechanisms leading to star formation both at large and small scales may be present at the same time. Surveys of the stellar clusters in the Rosette Molecular Cloud indicate that star formation in the region is primarily controlled by the primordial molecular cloud structure rather than the UV emission from the central OB cluster \citep{romanzunigaetal08, ybarraetal13}. However, \citet{wangetal10} suggest that the formation of some clusters was triggered by the expanding \ion{H}{ii} region and that the collect and collapse process \citep[reviews by e.g.,][]{elmegreen98, deharvengetal05} is forming new subclusters.

Various morphological features have been observed at the inner surface of these dusty molecular shells around OB associations, for example pillars (elephant trunks), and globules. Several mechanisms have been suggested to be behind their formation such as collect and collapse \citep{elmegreenlada77}, radiation-driven implosion (\citealt{reipurth83}; \citealt{bertoldi89}; \citealt{leflochlazareff94}), shadowing \citep{cerqueiraetal06, mackeylim10}, and collapse due to the shell curvature \citep{tremblinetal12}.
Triggered low-mass star formation has been observed also at these smaller scales \citep[e.g., cometary globule CG~1,][]{haikalaetal10, makelaetal13}.

The Rosette nebula (RN) surrounding the young cluster NGC~2244 is a well-studied \ion{H}{ii} region \citep[see e.g.,][and the references therein]{hbsfr_rosette}. \citet{lietal05} first identified a satellite cluster 6.6~pc west of NGC~2244, and \citet{wangetal10} studied the cluster that coincides with \object{NGC~2237} in closer detail. 
 \citet{romanzunigaetal08} used near-infrared (NIR) data to estimate the properties of the NIR excess stars over the region, and \citet{wangetal10} used \emph{Chandra} to search for X-ray emission from young low-mass stars. The large-scale distribution of molecular gas in the RN has been studied by \citet{schnepsetal80} and \citet{dentetal09}. The field has also been observed in the optical spectral region by \citet[][hereafter \citetalias{gahmetal07}]{gahmetal07}, who listed a number of tiny molecular clumps, globulettes, with a size distribution that peaks at $\sim$2.5~kAU and masses $\sim$1--700~M$_{\mathrm{Jup}}$.

We have previously examined elephant trunks, globulettes, and bright \Htwo\ rims in the northwest (NW) region of the RN in \citet[][hereafter \citetalias{gahmetal13}]{gahmetal13} and \citet[][hereafter \citetalias{makelaetal14}]{makelaetal14}. In \citetalias{gahmetal13} and \citetalias{makelaetal14} star formation was found to occur in the largest globulette and in one of the elephant trunks.

In this paper, we study the interaction between the \ion{H}{ii} region and the dusty molecular shell surrounding the central cluster in the western part of the RN. The region contains two dusty globules that protrude from the shell. The globules have sizes and masses that are an order of magnitude higher than the largest RN globulettes. These small globules at the rim of the shell are not as extended as elephant trunks nor are they separated from the shell like globulettes, but rather something in between.

We selected this field because one of the globules, named RN~A in \citetalias{gahmetal13}, has a curious morphology. Two slightly bent finger-like filaments extend from the main body of RN~A parallel to the shell. Along with the filaments, the core of RN~A gives the globule the shape of a seahorse. In addition, we report the details of the RN~A observations of \citetalias{gahmetal13} where a velocity gradient observed in the CO profile suggests a possible outflow. Earlier attention was drawn to this region by \citet{wangetal10} who found an X-ray source at the center of the globule, and by \citet{lietal05} who noted a highly reddened star next to the globule core. Some outflows are detected in the southern RN and the Rosette Molecular Cloud (e.g., \citealt{phelpsybarra05}; \citealt{dentetal09}) but so far only one is reported in the NW region of the RN, about 20\arcmin\ west of our observed field \citep{dentetal09}.

\citetalias{gahmetal13} suggest that globulettes form by detaching from the tips of elephant trunks. The larger globules at the rim of the shell may be precursors to free-floating globulettes and/or to the formation of more extended structures, like elephant trunks. In the present study we want to explore the nature of the two globules in the selected area and search for signs of on-going star formation. In particular, we focus the attention to globule RN~A with its striking finger-like extensions. From observations in radio molecular lines and NIR and MIR imaging we can conclude whether or not this morphology is a result of an outflow and if it can be related to the X-ray source or other IR stars found in this globule.

Observations and data reduction are described in Sect. \ref{sect:observations}, and the results are presented in Sect. \ref{sect:results}. Discussion is in Sect. \ref{sect:discussion} and a summary in Sect. \ref{sect:conclusions}. As we did in \citetalias{makelaetal14}, we adopt 1400~pc as the distance to the Rosette nebula \citep{oguraishida81}.


\section{Observations and data reduction}
\label{sect:observations}

During our previous studies of the region, RN~A was observed in \citetalias{gahmetal13} in \twco\ (1--0). We also include the Shell A position of that paper, which is located inside RN~A and an OFF position outside RN~A. The H$\alpha$ globulette survey by \citetalias{gahmetal07} covered this region in optical.

\subsection{NIR imaging}
\label{sect:imagingdata}

We used the Son of Isaac (SOFI) NIR camera on the New Technology Telescope (NTT) at the La Silla Observatory, Chile to observe a number of fields in the RN. The observations were done in jitter mode in the \jshks\ broadband filters and in the narrowband NB 1.282 P$\beta$, the continuum NB 2.09~$\mum$, and NB 2.124 \Htwo\ S(1) filters. The SOFI field of view is 4\farcm9 $\times$ 4\farcm9 with a pixel size of 0\farcs288. The observations were carried out in Jan. 2012. The NIR field discussed in the present paper is centered at $\alpha$=6:30:50.0, $\delta$= 5:02:10.0 (J2000.0). The other observed fields were discussed in \citetalias{makelaetal14}. We used the standard jittering mode with a jitter box width of 30\arcsec. The total integration time for the \Js\ and \Ks\ filters is 20 min and for the others 30 min. We observed standard stars from the faint NIR standard catalog by \citet{perssoncatalog} before and after the \jshks\ observations. The seeing during the observations was 0\farcs8--1\farcs6.

For data reduction, we used the IRAF external package XDIMSUM. We used the bad pixel, flat field, and illumination correction files on the ESO SOFI website for the broadband filters. For the narrowband filters, we used the flat field and illumination correction files taken during the observations. All frames went through crosstalk removal, bad pixel masking, cosmic ray removal, and stellar object masking. We removed the sky using two temporally closest frames to estimate the sky brightness. We applied the flat fields and illumination corrections and coadded the frames. We tied the image coordinate system to the 2MASS Point Source Catalogue (PSC).

Source Extractor (SExtractor) v. 2.5.0 \citep{SExtractor} was used to extract the \jshks\ photometry of the imaged field. It fits an aperture for each object and computes the flux inside the aperture. Zeropoints of the SOFI instrumental \jshks\ magnitudes were derived from the standard star observations. An initial photometry catalog was compiled from objects that had a detection and positive flux values in all bands. 

The extracted SOFI instrumental magnitudes were first converted into the Persson system \citep{perssoncatalog} and further into 2MASS magnitudes using the transformation formulae in \citet{ascensoetal07}. The difference between the \J\ and \Js\ filters was not taken into account. The transformation to the 2MASS magnitudes  was made to allow easy comparison with results reported in the literature and because the measurements that SExtractor flagged as saturated and that had 2MASS magnitudes <13\umag\ were replaced with 2MASS PSC data. A set of selection filters were used to remove the non-stellar objects and objects with non-reliable photometry from the SOFI measurements. These filters remove objects closer than 30~px of the image border; objects that had in any band a SExtractor flag other than 0 or 2 (i.e., the data were incomplete, pixel(s) were saturated, or a bright neighboring object influenced the photometry), a magnitude error larger than 0.15\umag, an extracted aperture elongation larger than 1.3, or a SExtractor stellar index less than 0.9 in more than one band. Objects with more than one star in the aperture were determined visually and also removed from the catalog.
The limiting \jshks\ magnitudes for a 0.15\umag\ error in the initial SOFI catalog are approximately 21.0 in \Js, 20.5 in \H, and 19.5 in \Ks. An OFF field created with the same filters in \citetalias{makelaetal14} was used as the reference field. The number of objects observed in the final catalog is 257 in the program (ON) and 389 in the OFF field. The \jhks\ photometry agrees well with the 2MASS PSC values. The SOFI \jhks\ photometry for the ON region is available as Table 1 at the CDS and for the OFF region photometry see \citetalias{makelaetal14}.

\addtocounter{table}{1}

\subsection{Radio observations}
\label{sect:apex}

We observed four positions in RN~A in the \twco\ and \thco\ (2--1) and (3--2) lines in the fall of 2012 with the 12~m APEX telescope at Llano Chajnantor, Chile. We carried out the observations in total power mode using the two single sideband heterodyne SIS-receivers mounted on the Nasmyth-A focus, APEX-1 and APEX-2. The APEX 230~GHz and 345~GHz full width at half maximum (FWHM) beam sizes are 27\arcsec\ and 18\arcsec\ and the main beam efficiencies 0.75 and 0.73. We used the RPG eXtended bandwidth Fast Fourier Transform Spectrometer that has 32\,768 channels and a bandwidth of 2.5~GHz. The channel spacing is 76.3~kHz, which corresponds to $\Delta v$ of 0.1 and 0.07~km\,s$^{-1}$ at 230 and 345~GHz, respectively. The pointing was checked regularly and the error is estimated to be better than 2\arcsec.

The effective system temperature when reduced to outside of atmosphere was around 250~K and 165~K in the 345 and the 230~GHz bands, respectively. A second order baseline was subtracted from individual spectra. The final rms is about 0.1~K and 0.06~K in the 345 and 230~GHz band, respectively. The APEX spectra are available online at the CDS in FITS format.


\section{Results} \label{sect:results}

\subsection{H$\alpha$ and SOFI Imaging}

\begin{figure*}
\centering
\includegraphics [width=14cm]{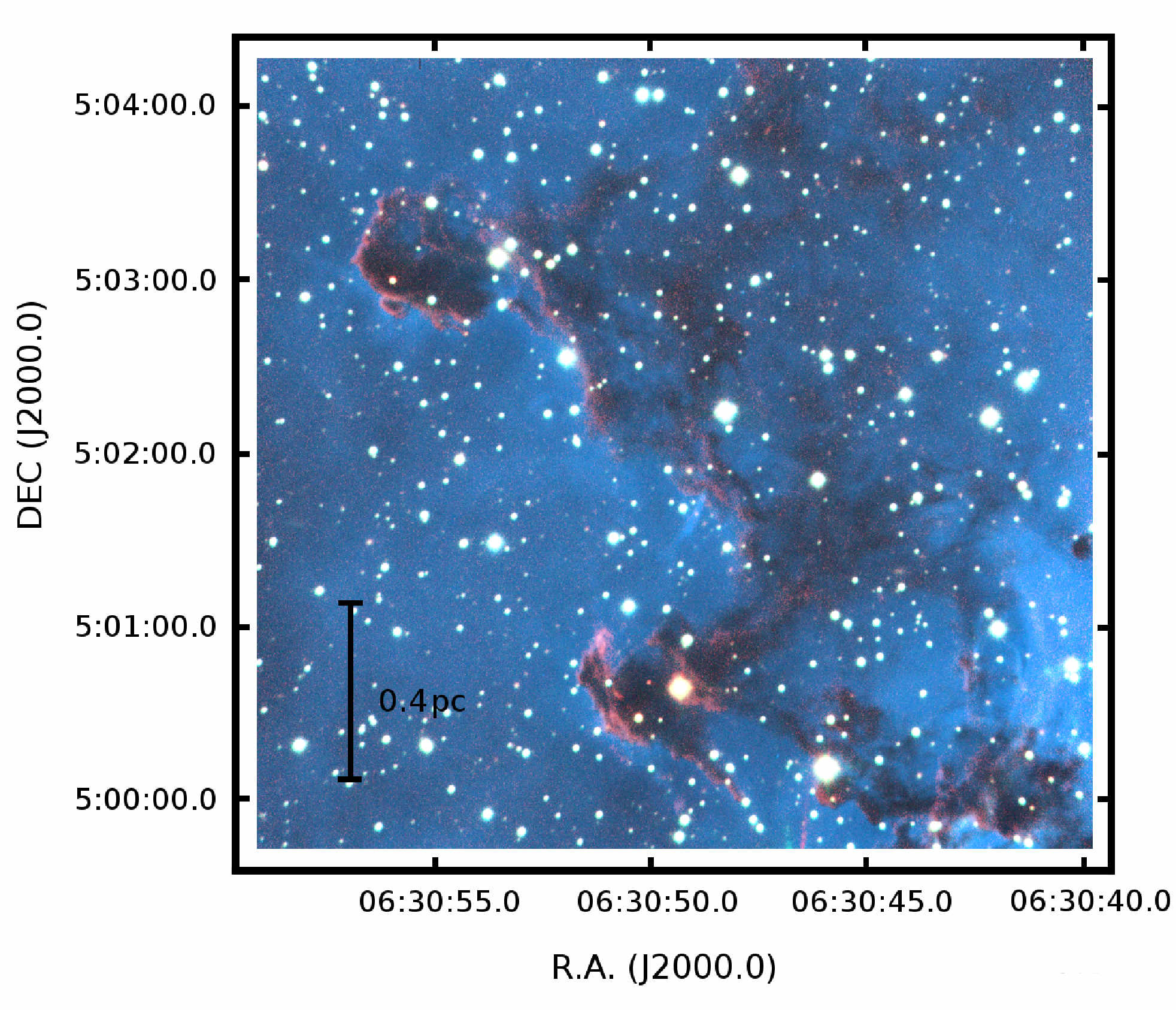}
   \caption{False-color image of the imaged SOFI field. H$\alpha$ and P$\beta$ are coded in blue, \H\ in green, and \Ks\ and \Htwo\ in red. Globule RN~A is below the center in the image, and globule RN~E is located at the top left corner.}
\label{fig:s3jshks}
\end{figure*}

\begin{figure}
\resizebox{\hsize}{!}{\includegraphics{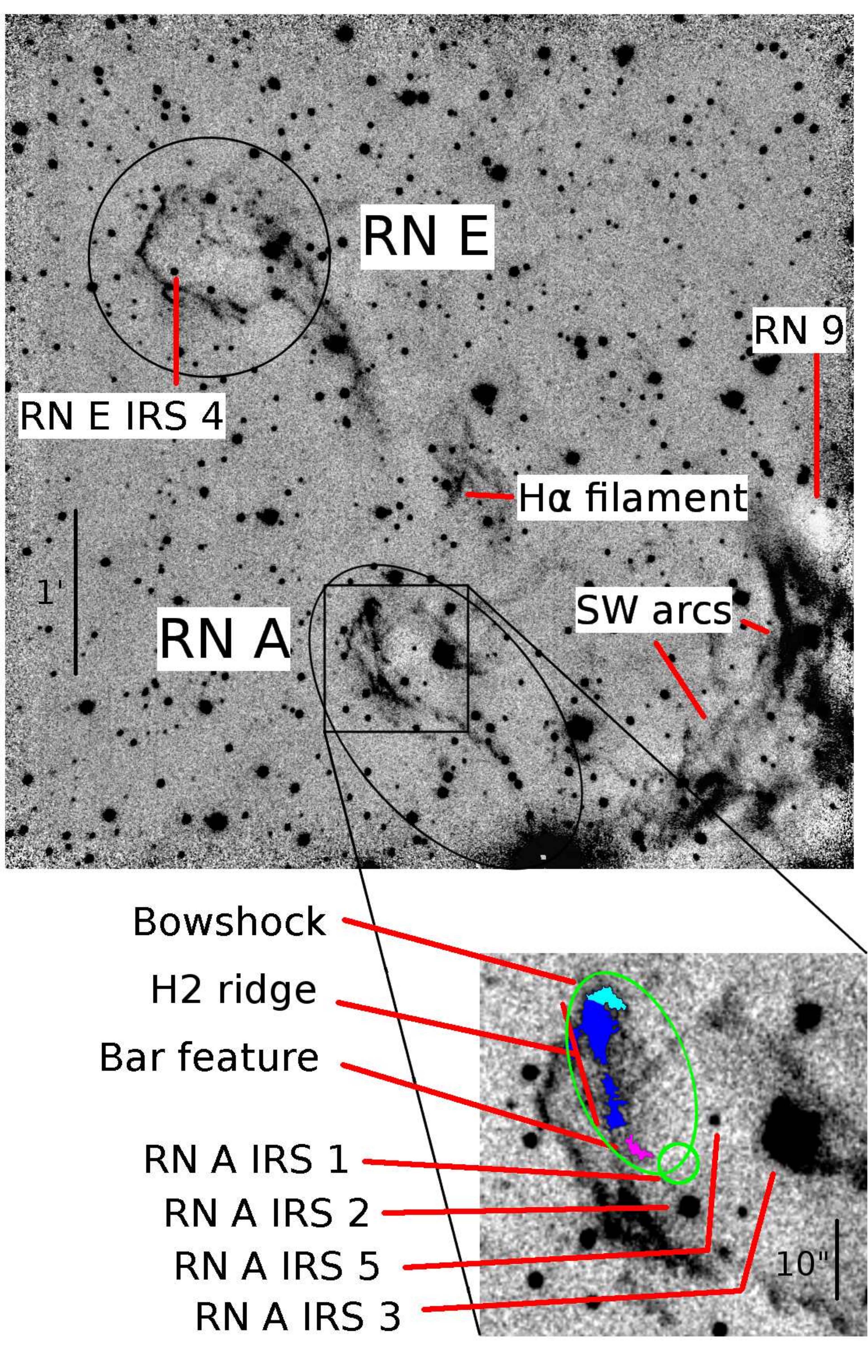}}
   \caption{Field of Fig. \ref{fig:s3jshks} presented as a combination of the \Htwo\ 2.12~$\mum$ and P$\beta$ images. The lower panel shows RN~A (``the Seahorse'') in closer detail. Some of the features discussed in the text have been marked.}
\label{fig:s3sketch}
\end{figure}

\begin{figure*}
\centering
\includegraphics [width=15cm]{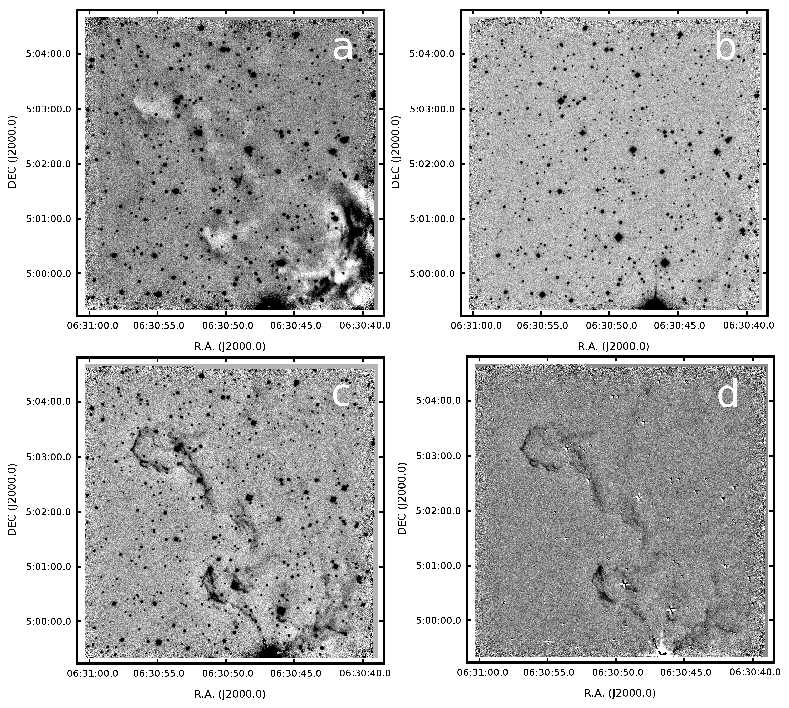}
   \caption{Grayscale images of the SOFI field in different wavelengths. a) P$\beta$ filter at 1.282~$\mum$. b) \Htwo\ continuum filter at 2.09~$\mum$. c) \Htwo\ 1--0  S(1) filter at 2.124~$\mum$. d) Negative difference image NB 2.124 \Htwo\ S(1)$-$2.09~$\mum$.}
\label{fig:s3nbbw}
\end{figure*}

A false-color image of the SOFI field combined from the NOT H$\alpha$ from \citetalias{gahmetal07} and SOFI P$\beta$, \H, \Ks, and \Htwo\ 2.12~$\mum$ images is shown in Fig. \ref{fig:s3jshks}. The individual SOFI grayscale images are shown in Figs. \ref{fig:s3jsbw}-\ref{fig:s3ksbw} and the NOT H$\alpha$ image in Fig. \ref{fig:s3halpha}. The features discussed in this paper have been marked in Fig. \ref{fig:s3sketch}. The observed narrowband images in P$\beta$, 2.09~$\mum$ continuum and \Htwo\ 1--0 S(1) 2.12~$\mum$ as well as the continuum-subtracted \Htwo\ image are shown in Fig. \ref{fig:s3nbbw}. Fig. \ref{fig:s3jshks} reveals the clumpy and ragged structure of the RN shell edge. The bright blue background is due to H$\alpha$ and P$\beta$ line emission. The dense material in the direction of the RN shell is seen in absorption against this bright background. Discrete bright H$\alpha$ filaments are seen in the center and in the lower right corner of the image. The bright red emission seen at the rims of the RN shell and the globules in Fig. \ref{fig:s3jshks} is due to \Htwo\ 2.12~$\mum$ line emission. Fainter emission is also seen on the surface of the shell.

Two globules protrude from the larger RN dust shell on the western side of the field. Because of the jittering observing mode, features larger than the jittering width (30\arcsec), such as the diffuse structure of the shell, are smeared out and brightness gradients are enhanced. Rim brightening of the globules and the shell boundary between the globules is seen in the NIR images in all filters except for the P$\beta$ line where faint diffuse emission is seen off the globules and shell boundary and in the NB 2.09 continuum where very little surface brightness is detected. The bright rims face the central cluster of the RN, and they are especially bright and thin in the \Htwo\ filter. The same phenomenon has been observed in the rims north and NW of this region \citepalias{gahmetal13, makelaetal14} where fluorescent \Htwo\ emission causes the bright rims. 

Both dark globules contain stars that are visibly reddened. The northern globule has one and the southern several reddened stars along with several non-reddened stars. The arc-shaped feature in the southwest (SW) corner (``SW arcs'') also contains several reddened stars.

The RN~A globule is elongated in the direction perpendicular to the bright rim. It is shaped like a seahorse, with a tail pointing SW from the core and a short head pointing northeast (NE). This morphology is very distinct in H$\alpha$ (see e.g., Fig. \ref{fig:s3jshks}). In Fig. \ref{fig:s3jshks}, the brightest RN~A star in the belly of the seahorse is reddened and we will call it RN~A IRS~3. The IRS designations used in this paper refer to stars that have a large $\H-\Ks$ color index (see Sect. \ref{sect:photometry}). They are discussed further in Sect. \ref{sect:sf}. Their locations are marked in the inset of Fig. \ref{fig:s3sketch} along with other notable features we will discuss in the following. About 20\arcsec\ east of RN~A IRS~3 a bar-like feature (``the bar'') is seen in the globule core in all filters.
The bar is elongated in a direction parallel to the Seahorse tail. A nearly north-south-directed bright ridge follows the edge of the globule and extends north of the bar into the Seahorse's head and it is seen as a red spot in Fig. \ref{fig:s3jshks}. The ridge is seen in the \Htwo\ 2.12~$\mum$ emission line, but not in the 2.09~$\mum$ continuum image (see Fig. \ref{fig:s3nbbw}). At the northern tip the ridge curves into a bowshock that is seen against the bright rim of RN~A. The \Htwo\ ridge and bowshock have been added to the molecular hydrogen emission-line object catalog as \object{MHO~3142} \citep{davisetal10}.
Other reddened stars are also seen in the direction of RN~A. \citet{wangetal10} found an obscured weak X-ray source ACIS \#86,
hereafter RN~A IRS~2, and RN~A IRS~3 was noted by \citet{lietal05} in a NIR study. RN~A IRS~5, in Fig. \ref{fig:s3jshks} is located 11\arcsec\ east of RN~A IRS~3.

We name the northern globule RN~E according to the naming convention started in \citetalias{gahmetal13}. This globule has a rounder shape than RN~A and no notable features other than the bright rims and one reddened star close to the southern edge of the globule which we will call RN~E IRS~4.
 
A small jet-like feature 15\arcsec\ in length can be seen protruding from the shell $\sim$1\arcmin\ NW of RN~A in P$\beta$ in Figs. \ref{fig:s3nbbw} and \ref{fig:s3halpha} and it is also seen in the H$\alpha$ and \Js\ bands. \citet{wangetal10} first noted this feature and suggested that it is an Herbig-Haro (HH) jet with a source 30\arcsec\ NW of the jet. However, the NIR images do not reveal a suitable candidate to drive a jet. We observed similar P$\beta$-bright filaments in \citetalias{makelaetal14} in the NW region of the RN and suggest that the feature is actually a bright ionized filament situated behind the shell and partly obscured by it. The SW arcs in the H$\alpha$ image are similarly ionized filaments.

Globulette RN~9 \citepalias{gahmetal07} is seen in absorption in the P$\beta$ image (Fig. \ref{fig:s3nbbw}) at the western edge of the field just north of the SW arcs. The P$\beta$ absorption and a slightly brightened rim in the \Htwo\ 2.12~$\mum$ line on the edge facing the central cluster are similar to the globulettes studied in \citetalias{makelaetal14}.

In the 2.09~$\mum$ continuum image (Fig. \ref{fig:s3nbbw}) the only non-stellar features are the bar in RN~A and some faint traces of the SW arcs, which indicates that the bright rims observed in the \Ks\ and \Htwo\ images are comprised of line emission. An image showing the \Htwo\ 2.12$-$NB 2.09 difference is shown in panel D of Fig. \ref{fig:s3nbbw}. Some artefacts from the stars remain because of incomplete subtraction. The bright rims and the RN~A bowshock remain but the bar in RN~A does not.

\subsection{Radio observations}

\begin{figure*}
 \centering
 \includegraphics [width=15cm] {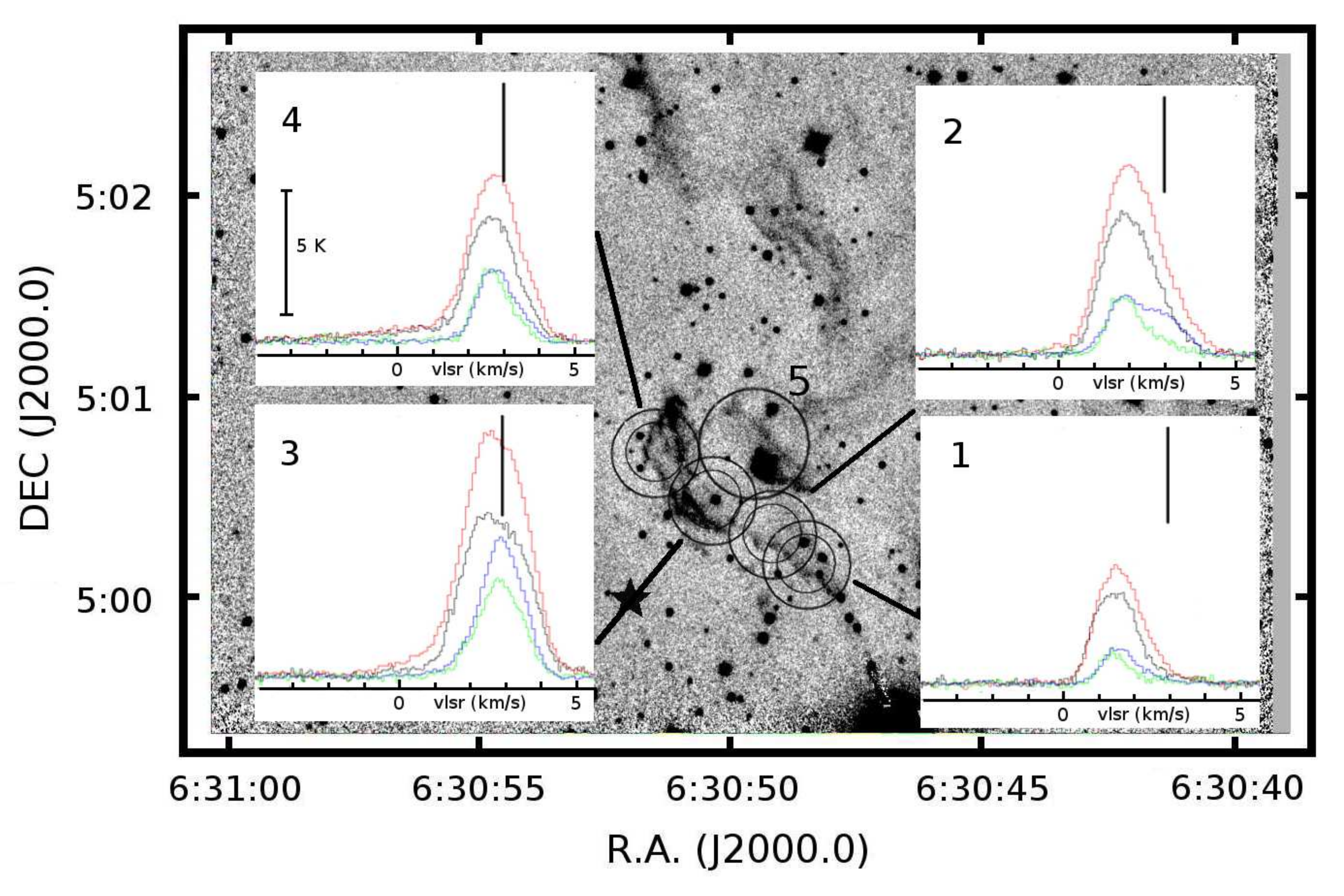}
    \caption{APEX beams and the observed spectra superposed on the RN~A \Htwo\ 2.12~$\mum$ image. The circle sizes correspond to the APEX FWHM at 230~GHz (27\arcsec) and 345~GHz (18\arcsec). The tickline in the spectra is at $v_{\mathrm{LSR}}=3$ \kmps. The \twco\ (3--2) and (2--1) are black and red, and \thco\ (3--2) and (2--1) lines are green and blue, respectively. The scale in the spectra is $T_{A}^{*}$. The circle with a thick line corresponds with the Shell A position of \citetalias{gahmetal13} and the star displays the OFF position.}
 \label{fig:apex_beams}
\end{figure*}

The four observed APEX positions in RN~A are marked in Fig. \ref{fig:apex_beams}. Additionally, one off-position where no signal was detected was observed southeast (SE) of RN~A. Positions 1 and 2 cover the Seahorse's tail. Position 3 is centered on the X-ray source RN~A IRS~2 in the main globule. Position 4 is the northernmost and there half of the APEX beam covers the main globule and a part of MHO~3142. An additional position NW of the globule core was observed with OSO in \twco\ by \citetalias{gahmetal13} (Shell A) and it is marked in Fig. \ref{fig:apex_beams}.

Two velocity components are detected in the APEX spectra. In the tail, position 1, the line is centered at LSR 1.4~\kmps\ whereas in the main globule, position 3, it is centered at 2.8~\kmps. 
In position 2, the APEX beam in the CO (2--1) transition traces both the tail and the main globule and both the 1.4 and 2.8~\kmps\ components are visible in the \thco\ spectrum.
 At position 5, the \twco\ (1--0) velocity component is broad and emission from both components is detected \citepalias[see][Fig. A.2]{gahmetal13}. The Onsala beam is however large (33\arcsec) and the main beam efficiency is only 30\% so it is difficult to tell where the observed emission is coming from. Only the 2.8~\kmps\ velocity component is detected in position 4 where the beam covers mainly MHO~3142.

The line intensity detected at position 3 is the largest of those observed but this may be because the beam is best centered on the globule of all the observed positions. Blueshifted line wings that are typical signs of a molecular outflow are observed in positions 3 and 4. No clear sign of redshifted wing emission is detected.

\subsection{Spitzer and WISE imaging}
\label{sect:iracwise}

\begin{figure}
\resizebox{\hsize}{!}{\includegraphics{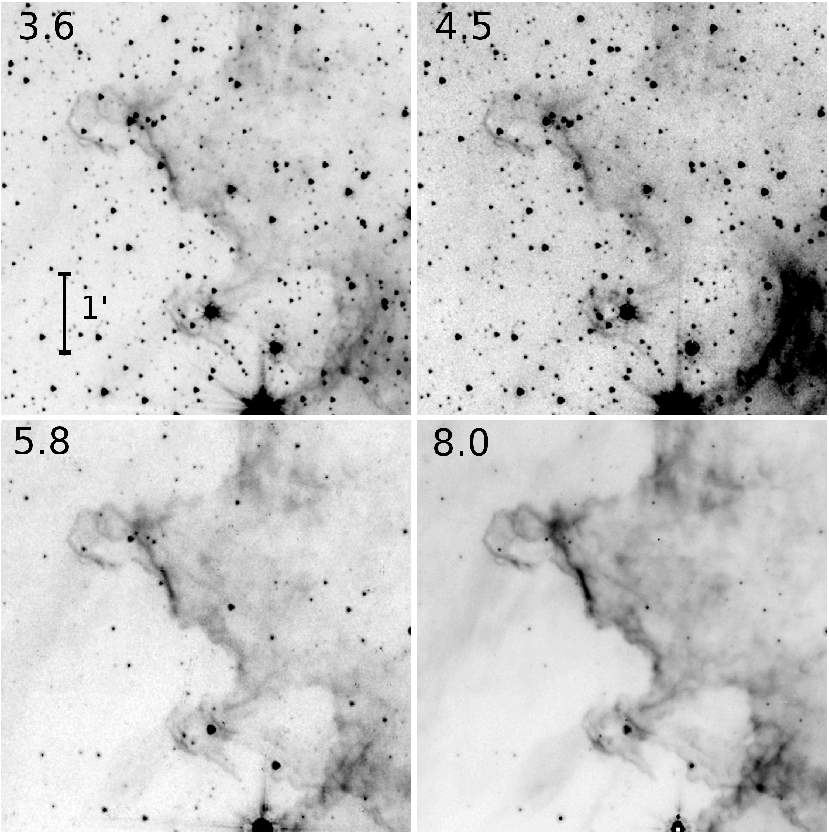}}
    \caption{\textit{Spitzer} IRAC images of the SOFI field (4\farcm9 $\times$ 4\farcm9) in negative grayscale. The IRAC wavelength is indicated in the upper-left hand corner in microns.}
 \label{fig:s3iracbw}
\end{figure}

We accessed the Spitzer Heritage Archive to acquire \textit{Spitzer} images of the field. The area covered by SOFI has been imaged in the Infrared Array Camera (IRAC) bands (AORKEY 21923072/G. Rieke) but not in the Multiband Imaging Photometer for Spitzer (MIPS) bands. The IRAC bands are centered at 3.6, 4.5, 5.8, and 8.0~$\mum$. The images are shown in Fig. \ref{fig:s3iracbw}. A false-color image in the 3.6, 4.5, and 8.0~$\mum$ bands is shown in Fig. \ref{fig:s3iracrgb}. The IRAC images can be used to diagnose the origin of the observed emission (e.g., \citealt{smithetal06}; \citealt{qiuetal08}). Scattered light is seen in the 3.6 and 4.5~$\mum$ bands, and shocked \Htwo\ especially in the 4.5~$\mum$ band \citep{smithrosen05}. The 4.5~$\mum$ filter contains also the Br$\alpha$ (4.05~$\mum$) hydrogen recombination line and several \Htwo\ lines. Polycyclic aromatic hydrocarbon (PAH) particles emit in all IRAC bands except at 4.5~$\mum$.

The molecular shell and the dark globules are seen in all bands through surface brightness and the morphology is the same as in the SOFI \jshks\ images. Bright rims are seen in all bands and an especially bright rim is located SW of RN~E. The SW arcs dominate the 4.5~$\mum$ image. At 5.8 and 8.0~$\mum$ the arcs are fainter and the surface structure of the cloud is seen better along with the dusty lane perpendicular to the arcs. The head of globulette RN~9 has a bright rim at 3.6~$\mum$.

The bar in RN~A is seen in all IRAC bands, and its shape changes from elongated to point-like as the wavelength increases so that at 8.0~$\mum$ it is seen as a bright point source. This point source is located $\sim$1\arcsec\ SW from the southern tip of the \Ks\ bar.
In the Seahorse's head, MHO~3142 is seen as a bright feature at 4.5~$\mum$, and faintly at 3.6~$\mum$ but not at the longer wavelengths. Based on the location of the bowshock, the bar, and the blueshifted CO line wings, we assume in the following that RN~A has a nearly north-south-directed outflow that is powered by the 8.0~$\mum$ \textit{Spitzer} source, which we will call RN~A IRS~1. This makes MHO~3142 a part of the outflow cavity wall and the bowshock. Several stars are seen in the 8.0~$\mum$ image and some of them have been classified in earlier surveys (\citealt{romanzunigaetal08}; \citealt{wangetal10}; \citealt{cambresyetal13}). RN~A IRS~2 and IRS~3 can be seen in the 8.0~$\mum$ image along with the highly reddened star RN~E IRS~4.

Wide-field Infrared Survey Explorer (WISE) \citep{wrightetal10} images were obtained from IPAC through the WISE Image Service to supplement the \textit{Spitzer} observations at longer wavelengths. PAH emission will be observable in the 12~$\mum$ image and thermal emission from very small grains (VSGs) in the 22~$\mum$ image. The 12 and 22~$\mu$m images are shown in Fig. \ref{fig:s3_wise}. A false-color image at 3.4, 4.6, and 22~$\mum$ is shown in Fig. \ref{fig:s3wisergb}. The emission features seen in the WISE bands 3 and 4 (at 12 and 22~$\mum$, respectively) are similar, and both globules RN~A and RN~E are faint relative to the bright shell edge between them. The point source RN~A IRS~3 is visible but not for example RN~A IRS~1. Some emission is seen in the direction of the bright \Htwo\ rim at the SE edge of RN~A especially at 12~$\mum$. The SW arcs are also bright in the 12 and 22~$\mum$ bands, and they are seen in the SOFI NIR and IRAC 3.4~$\mum$ bands likely due to scattered light.

\subsection{Photometry}
\label{sect:photometry}

\begin{figure}
\resizebox{\hsize}{!}{\includegraphics{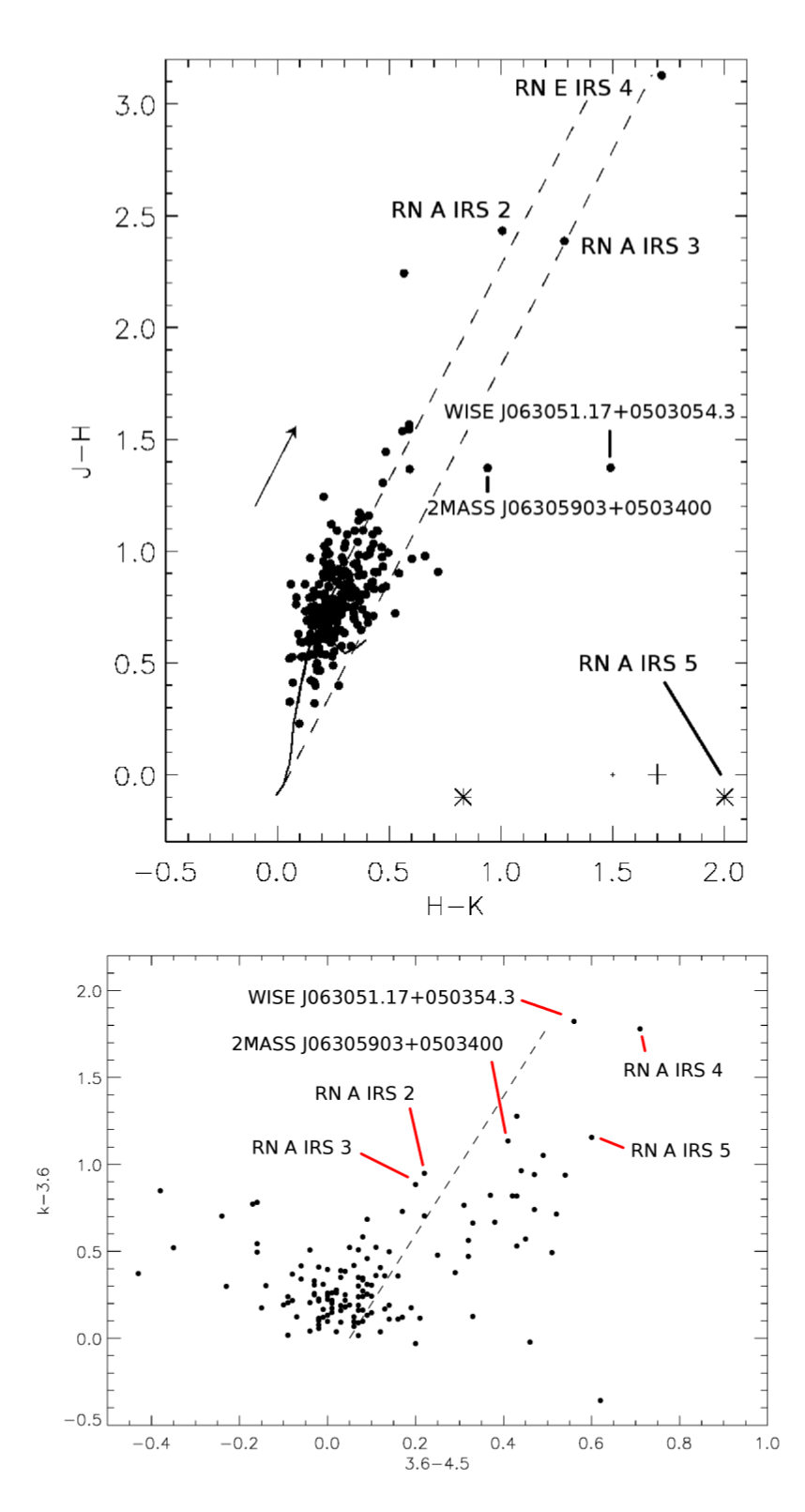}}
   \caption{\textit{Top:} $\J-\H$, $\H-\Ks$ color-color diagram of the objects in the observed SOFI field. Filled circles mark the final photometry catalog stars. Seven stars have been added from the 2MASS PSC catalog as their SOFI data were saturated. The dashed lines are the reddening lines, and the arrow is a 3\umag\ reddening vector. The \citetalias{bessellbrett88} reddening law has been used. The crosses mark stars detected only in \H\ and \Ks. The plus signs in the lower righthand corner mark the sizes of the average (left) and the maximum (right) error of the colors.
   \textit{Bottom:} \Ks--[3.6] vs. [3.6]--[4.5] color-color diagram. The dashed line is in the direction of the reddening vector \citep[adapted from][]{hartmannetal05, winstonetal07}, and IR excess objects lie right of the line.}
\label{fig:jhhkdiagram}
\end{figure}

The objects from the final SOFI photometry catalog have been plotted into a $\J-\H$, $\H-\Ks$ diagram in Fig. \ref{fig:jhhkdiagram} where some objects are labeled. Another seven stars were added to the catalog by using 2MASS PSC data to replace saturated SOFI data. Saturation takes place in SOFI photometry at magnitudes brighter than 14.5\umag. The extinction law used to draw the reddening line in Fig. \ref{fig:jhhkdiagram} is from \citet[][hereafter \citetalias{bessellbrett88}]{bessellbrett88}.
Most of the observed stars have been reddened by at least $\sim$1--2\umag. The highest reddening is observed for the star RN~E IRS~4 close to the rim in the northern globule RN~E. The star is discussed further in Sect. \ref{sect:rne}.

The two stars at $\J-\H\sim2.4$ are RN~A IRS objects 2 and 3, which do not exhibit large NIR excesses in Fig. \ref{fig:jhhkdiagram}. For RN~A IRS~2 \citet{wangetal10} estimated an visual extinction ($A_{V}$) of 20\umag, and for RN~A IRS~3 \citet{lietal05} estimated $\sim$17\umag. 
At $\J-\H\sim1.4$ two stars lie far below the lower reddening line where NIR excess stars are situated. The star \object{WISE J063051.17+050354.3} is located 91\arcsec\ NW and the star \object{2MASS J06305903+0503400} lies 60\arcsec\ NE of RN~E IRS~4.

The star at $\J-\H=2.2$, $\H-\Ks=0.6$, high above the upper reddening line, appears stellar in the \jshks\ image. Its position in the $\J-\H$, $\H-\Ks$ diagram may result from the very uneven background close to the star as it is located in the region of the SW arcs and dust lanes. Other possible explanations for such outliers are covered in \citet{fosteretal08}. 
Five other stars below the lower reddening line are located toward the RN dust shell except for one. Two of these have been labeled as NIR excess stars by \citet{romanzunigaetal08}.

Two stars that have no \Js\ detection but satisfy the other selection rules mentioned in Sect. \ref{sect:imagingdata} have been added as crosses in Fig. \ref{fig:jhhkdiagram}. We use the method of \citet{suutarinenetal13} to transform their instrumental SOFI $\H-\Ks$ color to 2MASS color. The star at $\H-\Ks=2.0$ is RN~A IRS~5 near the center of the RN~A core, which supports the large $\H-\Ks$ color. If we assume that it has no NIR excess, it has an $A_{V}\sim30$\umag\ which we will use as the ultimate upper limit for the RN~A $A_{V}$ value. The second star without a \Js\ band detection is located in the region where the SW arcs are intersected by a dust lane.

The \textit{Spitzer} IRAC magnitudes for these stars have been listed in \citet{kuhnetal13} as a part of the extensive MYStIX X-ray and infrared survey \citep{feigelsonetal13}. To get a more comprehensive view of the stellar population in the field, we matched the SOFI \jhks\ catalog objects with the MYStIX IRAC catalog. The resulting SOFI-IRAC catalog has 139 sources. Most of the MYStIX catalog objects have measured magnitudes only at 3.6 and 4.5~$\mum$, and therefore we have plotted the objects in a $\Ks-[3.6]$ vs. $[3.6]-[4.5]$ diagram. The diagram is shown in Fig. \ref{fig:jhhkdiagram} along with a line that separates the IR excess objects from reddened stars \citep{hartmannetal05, winstonetal07}. There are in total over 40 objects with apparent IR excess, and about 1/4 of them are located behind the dust shell structures. The rest are on lines of sight not obscured by the dust shell with most of them located east of the shell (i.e., on the cluster side). The IRS sources RN~A IRS~4 and 5 display IR excess in Fig. \ref{fig:jhhkdiagram}, but RN~A IRS~2 and 3 do not. RN~A IRS~1 is not plotted as it has no \Ks\ magnitude, but it has an IRAC $[3.6]-[4.5]$ color of 1.5, suggesting either IR excess or high reddening if not both. We also used the IRAC colors of the IRS sources to estimate whether they are young stellar objects (YSOs). Further confirmation for the classification was done by using their WISE magnitudes and the YSO class of RN~A IRS~2 and 3 has also been previously studied in the literature.

\subsection{Surface brightness}
\label{sect:sb}

We computed the surface brightness in the globulette and shell rims to evaluate their excitation mechanism. Typically, a line ratio is used to probe the excitation conditions. However, we observed in broadband filters and in only one spectral line that is typically used in such studies, that is, \Htwo\ 1--0 S(1).
We determined the surface brightness ratio between the \Htwo\ 1--0 S(1) line and \Ks, which enabled us to evaluate qualitatively the possible excitation mechanism. The \Ks\ filter covers the wavelengths of several \Htwo\ lines and the relative \Htwo\ line intensities depend on the ambient conditions. For example, at low temperatures the higher \Htwo\ energy levels are not populated and most of the \Htwo\ lines contribution in the \Ks\ filter comes from the 1--0 S(1) line.
	
We selected 17 positions on the bright rims and the outflow cavity walls and averaged the flux to represent the rim. The process is the same as described in \citetalias{makelaetal14}. The measured surface brightness values in the separate filters are of the same order as in the brighter rims in \citetalias{makelaetal14}. The \Htwo/\Ks\  ratios indicate that about about 20--40\% of the emission in the \Ks\ band is due to the \Htwo\ 2.12~$\mum$ line. This agrees with the \citet{blackvandishoeck87} values and is consistent with \citetalias{gahmetal13} and \citetalias{makelaetal14}, suggesting the same emission mechanism, that is, fluorescent \Htwo.
The only exceptions are the bar and the bright ionized filament classified by \citet{wangetal10} as a HH jet where the ratio is lower ($\sim$0.10). The \Htwo\ 2.12~$\mum$ contribution relative to \Ks\ in them is smaller, which is consistent with the view that the surface brightness in the bar is caused by scattered light rather than \Htwo\ emission. The observed 2.09~$\mum$ continuum surface brightness in the bar further supports this.

%

\section{Discussion} 
\label{sect:discussion}

\subsection{Visual extinction}
\label{sect:nicermap}

We use the NICER method \citep{nicer_lombardietalves} to estimate the large scale distribution of $A_{V}$ in the imaged region. In the areas with a small number of observed stars, such as the globules, the statistical accuracy will be low. In dense regions the resulting $A_{V}$ map values skew toward lower values both because we do not observe the highly obscured faint stars and because the $A_{V}$ determination may use stars outside the dense areas. The stars that have clear NIR excess (named in Fig. \ref{fig:jhhkdiagram}) were not used when estimating the extinction because the NIR excess can cause bias in the $A_{V}$ values. We will study the IRS objects in closer detail using individual line-of-sight extinctions in Sect. \ref{sect:pillarav} to estimate the extinction caused by the globules. The NICER method can use stars that have photometry in at least two bands to create an $A_{V}$ map, so we use as an input the SOFI ON catalog which is augmented with stars discussed in Sect. \ref{sect:photometry}, and an OFF field observed in \citetalias{makelaetal14}. Some of the stars in the NICER catalog will be foreground stars, but we have no method to reliably separate these from the unreddened stars at the distance of NGC~2237 and hence we did not attempt to exclude them. Including the foreground stars will bias the $A_{V}$ toward lower values around them. 

\begin{figure}
\resizebox{\hsize}{!}{\includegraphics{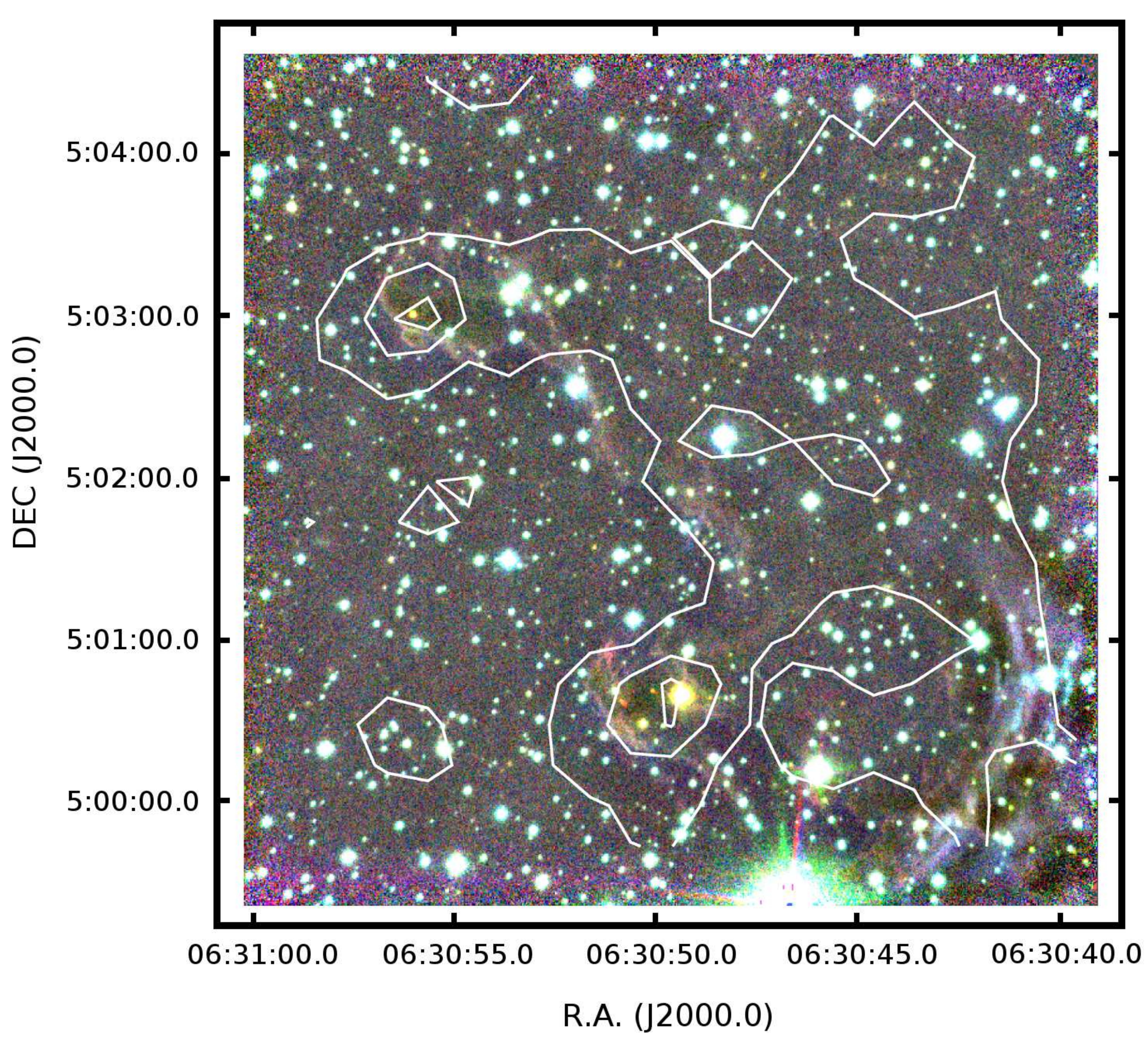}}
   \caption{$A_{V}$ contours on the false-color SOFI \jshks\ image (the \Ks, \H, and \Js\ bands are coded red, green, and blue, respectively). Contours 2, 5, and 8\umag\ are drawn.}
\label{fig:avmap}
\end{figure}

The resulting $A_{V}$ map is shown in Fig. \ref{fig:avmap}. It has a pixel size of 15\arcsec\ and the weighting Gaussian has a FWHM of 30\arcsec. The globules RN~A and RN~E have the highest $A_{V}$ values and are easily identified. The maximum $A_{V}$ values in the northern and southern globules are 8.6\umag\ and 9.4\umag, respectively. The dusty shell has $A_{V}$ values ranging from $\sim$2 to 3\umag. In the central region of the RN the $A_{V}$ is $\sim1-1.5$\umag.

\subsection{Star formation in the globules}
\label{sect:sf}

The IRS objects identified in the globules are listed in Table \ref{table:irstars}. We study these objects more closely to determine whether they are YSOs or highly reddened main sequence stars.
We have utilized the SED Fitting Tool \citep{robitailleetal07} to examine four of these in closer detail. We selected these stars as they are the only objects seen in the globules at 5.8 and 8.0~$\mum$ but not in H$\alpha$ suggesting they are not foreground stars. The online tool takes the stellar fluxes at different wavelengths as the input and compares the resulting spectral energy distributions (SED) to a precomputed model grid of stellar SEDs. Stellar magnitudes in WISE have been retrieved via IRSA, and \textit{Spitzer} IRAC magnitudes have been listed in the MYStIX catalog of \citet{kuhnetal13}.
 Also distance and interstellar $A_{V}$ ranges are provided before the fitting. We allowed a distance range of 1.2 to 1.7~kpc and the $A_{V}$ range was provided separately for each star based on the estimates in Sect. \ref{sect:photometry}.
We use the ``good'' fits where $\chi^{2}_{i}-\chi^{2}_{best}<3$ to derive the fitting results.

\begin{table*}
 \caption{Notable observed infrared stars in globules RN~A and RN~E.}
 \label{table:irstars}
 \centering
 \begin{tabular}{c c c c c l}
 \hline\hline
   ID & 2MASS designation & $\alpha$ & $\delta$ & YSO class & Comments   \\
 \hline
   RN A IRS 1 &      \ldots    			         & 06:30:50.74 & 05:00:34.9  & I & Outflow source \\
   RN A IRS 2 & \object{2MASS J06305033+0500288} & 06:30:50.33 & 05:00:28.85 & I/II & ACIS \#86 \\
   RN A IRS 3 & \object{2MASS J06304936+0500394} & 06:30:49.37 & 05:00:39.45 & II & Brightest RN~A star \\
   RN E IRS 4 & \object{2MASS J06305605+0503003} & 06:30:56.05 & 05:03:00.33 & I/II & \\
   RN A IRS 5 &      \ldots 					 & 06:30:50.07 & 05:00:41.4  & \ldots & No J magnitude\\
 \hline
 \end{tabular}
 \tablefoot{Column 1 states the designation we use in this paper, and Column 2 the 2MASS identifier. Columns 3 and 4 contain the J2000 coordinates, Column 5 the estimated YSO class, and Column 6 has additional comments.}
 \end{table*}

\begin{figure}
\resizebox{\hsize}{!}{\includegraphics{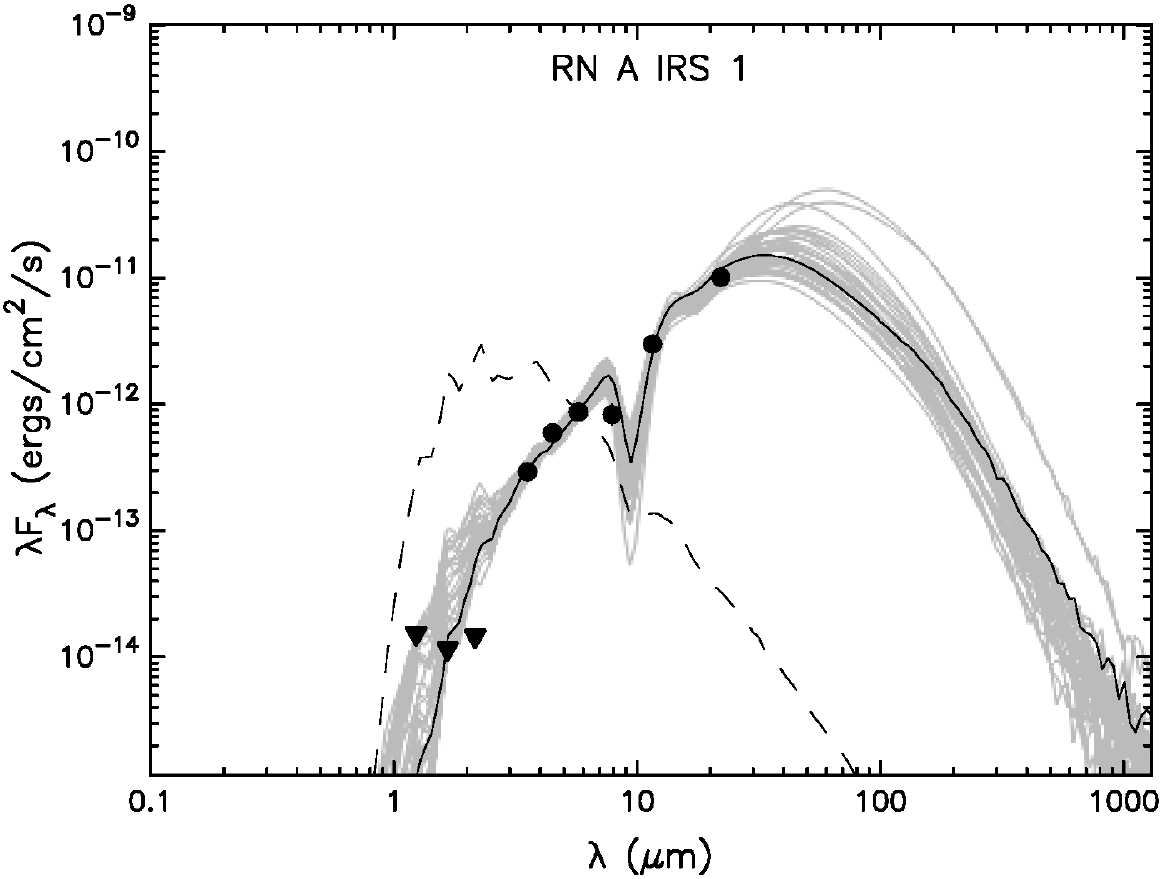}}
   \caption{SED for RN~A IRS~1. The triangles show the SOFI \jhks\ limiting magnitudes as upper limits. The circles show the flux values for IRAC and the WISE 12 and 22~$\mum$ fluxes. The best fit is shown with a solid black line, and fits with $\chi_{i}^{2}-\chi_{best}^{2} < 3$ are shown with gray lines. The dashed line is the stellar photosphere of the best fit including the reddening by interstellar extinction.}
\label{fig:sedfit_outflowstar}
\end{figure}

The outflow source candidate RN~A IRS~1 is not seen in our SOFI images, allowing only upper limit flux estimates in the NIR bands. We used the limiting magnitudes as upper limits and additionally set the confidence at 99\%. A SED fit for RN~A IRS~1 is shown in Fig. \ref{fig:sedfit_outflowstar}. This includes the \jhks\ upper limits, IRAC, and WISE 12 and 22~$\mum$ magnitudes. The allowed $A_{V}$ range is 5 to 25\umag. There are no data points for $\lambda$ > 22 $\mum$, which causes some uncertainty in the fits, but the WISE 12 and 22~$\mum$ data points indicate the existence of a disk component. The weighted mass mean value is 0.20~\Msun, and the best fit value is 0.15~\Msun. These should be considered to be order of magnitude estimates. The SED shape suggests that the outflow source is a Class~I YSO due to the large disk component. The listed IRAC values yield colors $[3.6]-[4.5]=1.51$ and $[5.8]-[8.0]=0.93$, which also indicate Class~I according to the YSO classification by \citet{allenetal04} and \citet{megeathetal04}.

For RN~A IRS~2 we fit a SED based on the SOFI, WISE, and the IRAC 3.6 and 4.5~$\mum$ magnitudes as the 5.8 and 8.0~$\mum$ magnitudes are not listed in the MYStIX catalog. We use an $A_{V}$ range of 10 to 20\umag. The resulting fit is shown in Fig. \ref{fig:sedfit_acis86}. The mean and best fit masses are 1.3 and 1.6~\Msun\, respectively. The shape of the SED and the spectral index suggest a flat spectrum (Class I/II) YSO. In the literature \citet{cambresyetal13} classify it as a Class~I object, but \citet{wangetal10} use the NIR colors to classify it as a Class~III source with a mass of $\sim$1.5~\Msun, which fits the mass estimate we derive with the SED Fitting Tool.

2MASS, WISE, and IRAC data are available for RN~A IRS~3. We use an $A_{V}$ range of 10--30\umag. The SED is plotted in Fig. \ref{fig:sedfit_rnabrightest}, and the shape suggests a Class~II YSO. The resulting mean and best fit masses are 2.9 and 2.1~\Msun, respectively. The IRAC colors $[3.6]-[4.5]=0.2$ and $[5.8]-[8.0]=0.02$ place it among Class~III YSOs in the \citet{megeathetal04} classification but the WISE 22~$\mum$ observation suggests it still has a disk component and is therefore a Class~II YSO. \citet{cambresyetal13} support a Class~II determination for RN~A IRS~3 based on its spectral index.

The highly reddened RN~E IRS~4 has \jhks\ magnitudes derived from SOFI observations as well as IRAC, and WISE magnitudes. The $A_{V}$ was limited to a range between 5 and 30\umag. The fit SED is shown in Fig. \ref{fig:sedfit_npillar}. The SED shape and the spectral index indicate a flat-spectrum (Class I/II) YSO. The mass range is not well constrained, but as an indication of the possible values, the weighted mean value is 0.4~\Msun. Its IRAC colors are $[3.6]-[4.5]=0.71$ and $[5.8]-[8.0]=0.29$, which indicate a possibly reddened Class~II YSO. The influence of the $A_{V}$ can be seen in Fig. \ref{fig:sedfit_npillar} where the high-$A_{V}$ fits have smaller disk components.

The RN~A IRS~5 SED has only four data points at the shorter IR wavelengths, and it has not been fit because the fits are difficult to constrain without datapoints at $\lambda>4.6~\mum$. The SOFI-IRAC color-color diagram in Fig. \ref{fig:jhhkdiagram} shows that it has IR excess, but the high reddening toward this star also needs to be accounted for. Measurements at longer infrared (IR) wavelengths would be needed to determine its nature.

\subsubsection{Globule masses}
\label{sect:pillarav}

We have estimated the mass of both globules by approximating their dark \Js\ cores with ellipses which are fit to the core by eye.
The core sizes are 16.5\arcsec$\times$10.5\arcsec\ for RN~A, and 23.0\arcsec$\times$7.8\arcsec\ for RN~E. We assume that the line-of-sight depth of the globules is the average of the axes of the fit ellipses. We used the \citet{bohlinetal78} relation for the mean column density of molecular hydrogen, $N(\Htwo)$, and visual extinction: $N(\mathrm{\Htwo}) = 0.94 \times10^{21}~\mathrm{cm}^{-2}~\mathrm{mag}^{-1}$. A mass of 2.8~u per \Htwo\ molecule and constant density are assumed when computing the masses. For each IRS object a weighted $A_{V}$ average was computed from the interstellar $A_{V}$  values produced by the SED Fitting Tool, and a mean $A_{V}$ was then derived for each globule. The fit $A_{V}$ values for the IRS objects 1--3 in RN~A are 11.1--15.5\umag, and the derived mean $A_{V}$ for RN~A is 13.6\umag. For RN~E the $A_{V}$ is 16.0 \umag\ based on RN~E IRS~4.

The resulting mean number densities in the globules are $4.5-4.7\times10^{4}~\mathrm{cm}^{-3}$ and the globule masses are 9.5--11.6~\Msun\ with RN~E being denser and more massive because of the larger fit ellipse and $A_{V}$. The density of the globules is similar to the mean density inside the RN globulettes ($n(\mathrm{\Htwo})\sim10^{4}~\mathrm{cm}^{-3}$; \citetalias{makelaetal14}).
Above we assumed that the IRS objects are located in the center of the globule. Placing the stars right behind the globules instead of their center, produces the lower limits for the masses which are half of the presented estimates. In both globules the dark \Js\ core extends to the bright \Htwo\ rims, indicating that $A_{V}$ is high already at the surface of the globule on the central cluster side. However, the less dense matter trailing the high-density cores on the shell-side will also add a contribution to the total mass of the globule. 
Even though the $A_{V}$ is not high in the tail, the trailing matter covers an area comparable to the core. We estimate the mass error to be $\pm$50\%.

\subsubsection{Star formation efficiency}

The star formation efficiency (SFE) for both globules has been computed using the globule mass estimate and the weighted mean mass for the IRS objects derived from the good SED fits. For RN~E the SFE is $\sim$4\%, but in RN~A the situation is more complex. RN~A IRS~2 is likely a background star and IRS~3 is located outside the region where the globule mass is estimated. The SFE computed from RN~A IRS~1 alone is $\sim$2\%. It should be noted that these SFE numbers are rough estimates. The SFE for small globules is estimated to be $\sim$6\% \citep{yunclemens90} and for dense cores the maximum SFE is $\sim$30\% \citep{alvesetal07}. The SFEs in the globules are at the lower end of these values. However, we cannot be certain whether other IRS objects than RN~A IRS~1 are located inside the globules or behind them.

This study provides a detailed look at a small portion of the RN shell that is seen projected on the cluster NGC~2237 which has over 200 members and maybe even as many as 600 \citep{lietal05, wangetal10}. The distance to the cluster NGC~2237 is uncertain. \citet{bonattobica09} suggested that the distance of cluster NGC~2237 is 3.9~kpc, but this has been discussed in \citet{wangetal10} and typically a distance similar to the distance of NGC~2244 is preferred \citep[i.e., $\sim$1.4~kpc, e.g.,][]{wangetal10}. 
In \citetalias{makelaetal14} we found two YSOs in the Rosette globulettes and elephant trunks NE of the region studied in this paper. Based on these combined results we conclude that the contribution from these small structures at the NW edge of RN to the stellar content of the NGC~2237 cluster is not significant.

\subsection{Globule RN~A}
\label{sect:rna}

The southern globule RN~A is shaped like a seahorse. The globule forms the body of the Seahorse and the head and tail are thin dusty extensions that lie parallel to the shell. The globule is connected to the larger shell structure at the belly of the Seahorse. The two velocity components seen in the APEX CO (3--2) spectra indicate that the body of RN~A and the Seahorse tail are separate structures. The spatial resolution of the large scale CO (3--2) velocity map in \citet{dentetal09} is 20\arcsec\ and can be used to study the velocity structure in the area. Because the velocity resolution  of the \citet{dentetal09} map is only 0.45~\kmps\ small discrete jumps in the velocity like that observed between the RN~A body and the Seahorse tail will blend together. The velocity map reveals however a clear velocity difference between RN~A and the large shell structure west of it.
\citet{dentetal09} measure a LSR velocity of 2.5~\kmps\ for RN~A. The shell velocity has been measured as 1.3--1.9~\kmps\ in \citet{schnepsetal80} using CO (1--0) observations which agrees with the APEX observations. The difference in velocity is similar to that between the RN~A body and the Seahorse tail.


\begin{figure}
\resizebox{\hsize}{!}{\includegraphics{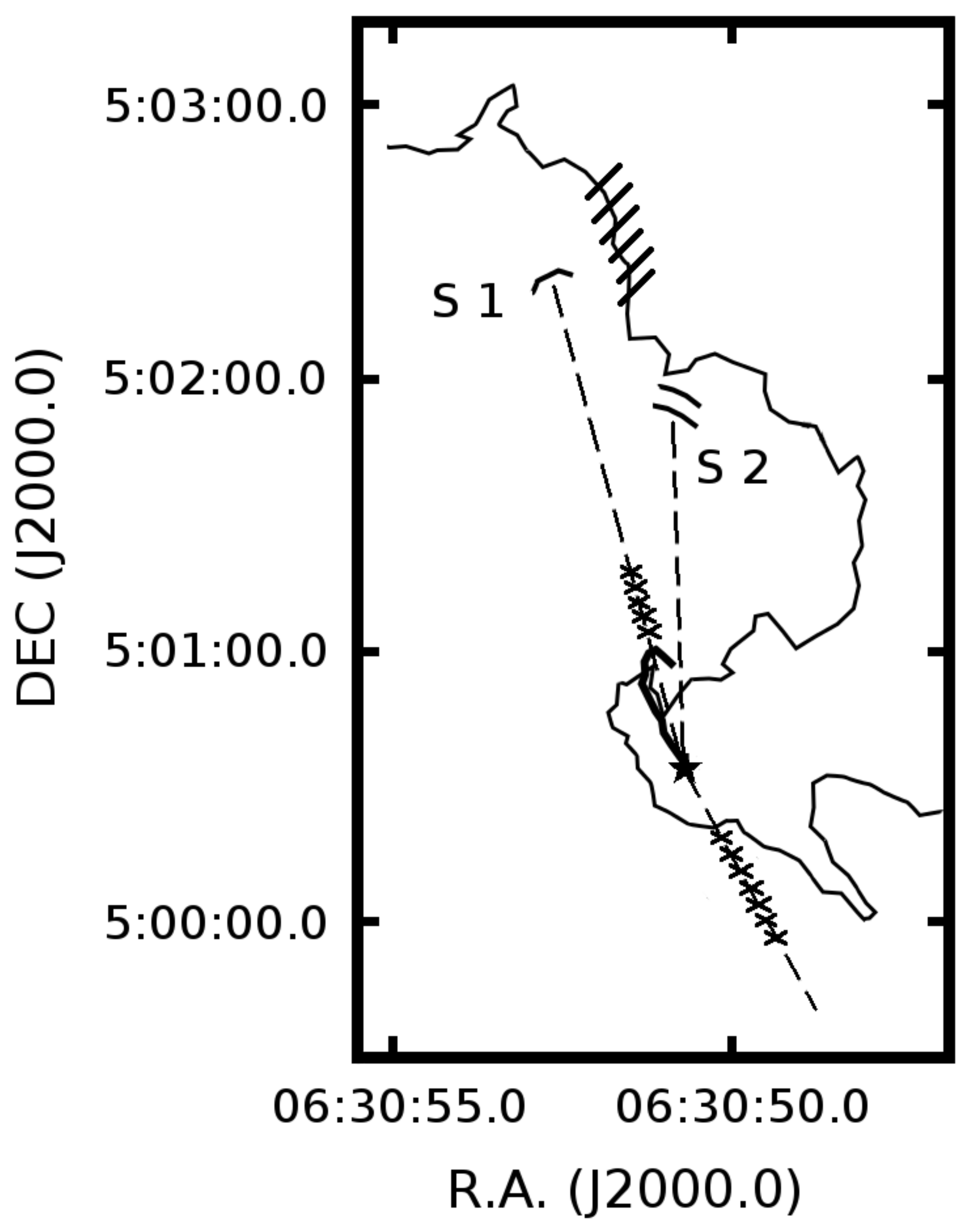}}
   \caption{Sketch of the RN~A outflow. The \Htwo\ outflow cavity is drawn with a thick solid line in RN~A and the outflow central source is marked with a star. The short arcs marked S~1 and S~2 display the positions of the H$\alpha$ bowshocks. The dashed lines have been drawn from the outflow source through the H$\alpha$ knots, which are marked with 'x' to indicate the direction of the possible outflow. The short lines on the shell display the section that is bright in the IRAC bands.}
\label{fig:rna_sketch}
\end{figure}

RN~A contains several obscured YSO candidates (see Sect. \ref{sect:sf}). In RN~A, IRS~1 is the youngest object, and IRS~2 and 3 are reddened older YSOs.
The outflow source RN~A IRS~1 shapes the view of the globule especially in the \Ks\ and \Htwo\ 2.12~$\mum$ bands. The bar inside the Seahorse and MHO~3142 are both a part of the eastern wall of the outflow cavity. See Fig. \ref{fig:rna_sketch} for a sketch of the globule. Especially the continuum detection of the bar at 2.09~$\mum$ indicates that it scatters light from the outflow source located at the southern end of the bar. Light scattering from outflow cavity walls has been previously observed by, for example, \citet{velusamyetal07} in HH~46/47. 
The ridge of MHO~3142 in turn is seen in \Htwo\ emission where the outflow meets the ambient medium. The \Htwo\ bowshock following the ridge is seen against the northern edge of the dust globule. This supports the view that the \Htwo\ emission comes from a cavity wall instead of a jet. A main-sequence star is located on top of the bright cavity wall, but it is most likely a foreground object as it is seen in the H$\alpha$ image.
\citetalias{gahmetal13} suggest that stars formed in the RN globulettes will move at high velocities away from the cluster. The LSR velocity of RN~A is close to that observed for globulettes in the neighboring regions \citepalias{gahmetal13} and differs from the RN system velocity by $\sim$20~\kmps. 

The outflow cavity is detected in the IRAC 4.5~$\mum$ image, which has no strong PAH features, and also faintly in the IRAC 3.6~$\mum$ image. This suggests it is shocked \Htwo\ \citep{smithrosen05, takamietal10}. The \Htwo\ 1--0 S(1) image confirms the outflow nature. The distance between the bowshock and the driving source is about 26\arcsec\ (roughly 0.18~pc at the distance of RN). 
The western cavity wall is not seen in the images. The region directly below the Seahorse's head is void of dust, which could indicate that the outflow has blown that region clean.

The blueshifted line wing in the \twco and \thco spectra in positions 3 and 4 reveals the northern outflow lobe. The line wing seems to broaden away from RN~A IRS~1, but our study has only five measured positions and more are needed to fully understand the RN~A complex. The redshifted line wing is not as clear as the blueshifted one, but there is an indication of modest redshifted emission in position 3. No redshifted emission is seen in the spectra of the Seahorse tail. The RMS of the \citet{dentetal09} CO data is too high to detect the low intensity outflow wings seen in the APEX spectra.

There is no clear indication of the southern outflow lobe in any of the observed IR bands. The northern lobe shows blueshifted line wings, suggesting that the southern lobe is pointed away from us and is more heavily obscured by RN~A. Small knots are seen south of RN~A in the H$\alpha$ image (Fig. \ref{fig:s3halpha}) parallel to the Seahorse tail. The \thco(2--1) beam in position 3 covers a small part of the knots closest to the globule, which may explain why only modest redshifted emission is seen in this position.

\subsubsection{A parsec-scale outflow?}

The features discussed below are sketched out in Fig. \ref{fig:rna_sketch}. The H$\alpha$ image in Fig. \ref{fig:s3halpha} shows weak emission at the location of the \Htwo\ bowshock. Several H$\alpha$ knots are seen starting from the RN~A bowshock and extending north for about 20\arcsec\ from the edge of the globule and for about 40\arcsec\ from the outflow source RN~A IRS~1. The WISE 12~$\mum$ surface brightness traces these knots. A line of faint H$\alpha$ knots may also trace the southern outflow arm to a distance of about 40\arcsec\ from RN~A IRS~1.
The driving source, \Htwo\ bowshock, H$\alpha$ knots, and the 12~$\mum$ surface brightness are all located roughly on the same axis. About 110\arcsec\ north of RN~A IRS~1 and east of the IRAC-bright shell is a faint H$\alpha$ feature, here called S~1, that is shaped like a bowshock. It is located on the same outflow axis. These H$\alpha$ knots and bowshocks could signal the progress of the outflow outside the matter-rich RN~A. At the distance of the RN, the farthermost H$\alpha$ bowshock would be 0.75 pc from the driving source.

Another pair of H$\alpha$ bowshocks, S~2, can be seen about 80\arcsec\ north of RN~A IRS~1 and west of the larger outflow axis. These are located below the shell rim that is peculiarly bright in the 5.8 and 8.0~$\mum$ IRAC images (Fig. \ref{fig:s3iracbw}). The shell is bright also at 3.6 and 4.5~$\mum$ where the latter rules out PAH emission. It is possible that the surface brightness is due to the outflow impacting the shell material.

\subsection{Globule RN~E}
\label{sect:rne}

RN~E is more symmetric of the two observed globules. The most reddened star in the $\J-\H$, $\H-\Ks$ diagram of our survey, RN~E IRS~4, lies in the head of the RN~E globule. RN~E IRS~4 is likely a reddened Class I/II star, and the RN~E images show no other signs linked with star formation. 
The X-ray study of \citet{wangetal10} lists one additional YSO candidate for RN~E at its northern edge. However, the star shows no NIR excess and is seen in the H$\alpha$ image, so it is likely a foreground star.

The velocity gradient for the RN~E globule in the radial direction to the RN central cluster was first determined in \citet{schnepsetal80} who observed it in \twco and \thco (1--0) and found a velocity gradient of $\sim$1.1~\kmps~arcmin$^{-1}$ (RN~E is their globule F). More recently, \citet{dentetal09} observed the Rosette in \thco\ (3--2) and measured a velocity gradient of 0.8~\kmps~arcmin$^{-1}$ and an LSR velocity of 2.4~\kmps\ for RN~E. The velocity gradient suggests that the globule is stretching with the tip closer to the central cluster moving slower and the end moving faster toward us.


\section{Summary and conclusions}
\label{sect:conclusions}

We have observed a region of the Rosette nebula in NIR \jshks\ filters and the globule RN~A in \twco\ and \thco\ (2--1) and (3--2) lines. We have supplemented the observations with archive WISE and \emph{Spitzer} IRAC data.
These are used to study the morphology and signs of star formation in the two semi-attached globules, RN~A and RN~E, and to determine whether RN~A harbors an outflow. We examine the ratio of \Htwo\ (1--0) S(1) 2.12~$\mum$ and \Ks\ surface brightness in 17 locations throughout the field, including bright \Htwo\ rims, an H$\alpha$ filament, and an outflow cavity. We constructed a visual extinction map using the \jhks\ photometry and probe the globule masses using individual stars in direction of the globules. All available photometric data were then combined to plot the SEDs for stars in or behind the globules.

In \citetalias{makelaetal14} and this paper we have studied star formation in the Rosette nebula at the interface between an \ion{H}{ii} region and a molecular shell. We have found only few cases of on-going star formation in the pillars and globules in the shell surrounding NGC~2244. Despite the detected star formation activity, the expanding \ion{H}{ii} region does not appear to have triggered wide-spread star formation and most of the active star formation in Rosette is taking place in the clusters.

\begin{itemize}
\item Two dark globules at the edge of the NW shell of the Rosette nebula, RN~A and RN~E, are detected in absorption in \Js\ and P$\beta$. They contain highly reddened stars with $A_{V}$ $\sim$14--16\umag. The globule masses are estimated to be $\sim$10--12~\Msun.

\item  On-going star formation has been observed in the direction of both globules. The single highly reddened star seen toward RN~E is likely a Class I/II YSO. In RN~A, two highly reddened YSOs (Class I/II and Class~II) are seen in the NIR and IRAC bands. The third star, RN~A IRS~1, is obscured in the NIR bands but detected in the 8.0~$\mum$ IRAC image. It is most likely a Class~I YSO embedded in RN~A. Submillimeter continuum observations would help to further constrain the ages and masses of the YSOs.

\item  RN~A IRS~1 drives an outflow. The CO spectra taken in two positions inside the globule body have a blueshifted component especially in the \twco\ lines that marks the northern outflow lobe. The eastern wall of the outflow cavity is bright in \Htwo\ 2.12~$\mum$ and at IRAC 4.5~$\mum$ emission. The base of the cavity wall is seen in scattered light in several NIR bands. The outflow cavity apex has a bowshock shape and lies at the northern edge of RN~A. The western cavity wall is not detected. The southern outflow lobe is not seen in the CO spectra taken in the Seahorse tail. Further molecular line observations of the outflow would be needed to explore its parameters such as velocity, direction, and extent.

\item H$\alpha$ knots are seen outside RN~A and they extend $\sim$40\arcsec\ from the outflow source RN~A IRS~1. A faint H$\alpha$ bowshock 110\arcsec\ north of RN~A is seen on the same axis as the knots. The distance between the farthest bowshock and the tip of the southern knots is 1~pc, suggesting that RN~A IRS~1 drives a parsec-scale outflow. However, additional observations would be needed to confirm the existence of the jet and also to probe its properties.

\item The \Htwo/\Ks\ ratio in the rims of the RN shell and globules is 20--40\%. This is similar to the globulettes in \citetalias{gahmetal13} and indicates that they are seen in fluorescent \Htwo\ emission. The ratio is lower in the base of the outflow cavity, which is also seen in the 2.09~$\mum$ continuum images and in the bright H$\alpha$ filament, which was previously in the literature considered to be an HH object. 

\end{itemize}

\begin{acknowledgements}
M.M. acknowledges the support from the Doctoral Programme in Astronomy and Space Physics, the Academy of Finland Grant No. 250741, and the DFG grant SA 2131/4-1. This research has made use of NASA's Astrophysics Data System, and the SIMBAD database, operated at CDS, Strasbourg, France. This work is based in part on observations made with the {\textit Spitzer Space Telescope}, which is operated by the Jet Propulsion Laboratory, California Institute of Technology under a contract with NASA. This publication makes use of data products from the Wide-field Infrared Survey Explorer, which is a joint project of the University of California, Los Angeles, and the Jet Propulsion Laboratory/California Institute of Technology, and NEOWISE, which is a project of the Jet Propulsion Laboratory/California Institute of Technology. WISE and NEOWISE are funded by the National Aeronautics and Space Administration. IRAF is distributed by the National Optical Astronomy Observatories, which are operated by the Association of Universities for Research in Astronomy, Inc., under cooperative agreement with the National Science Foundation. The MHO catalog is hosted by Liverpool John Moores University.
\end{acknowledgements}

\bibliographystyle{aa}
\bibliography{s3_references.bib}

\listofobjects


\begin{appendix}

\section{Additional figures}
 \label{app:figures}

\begin{figure*}
 \centering
 \includegraphics [width=17cm] {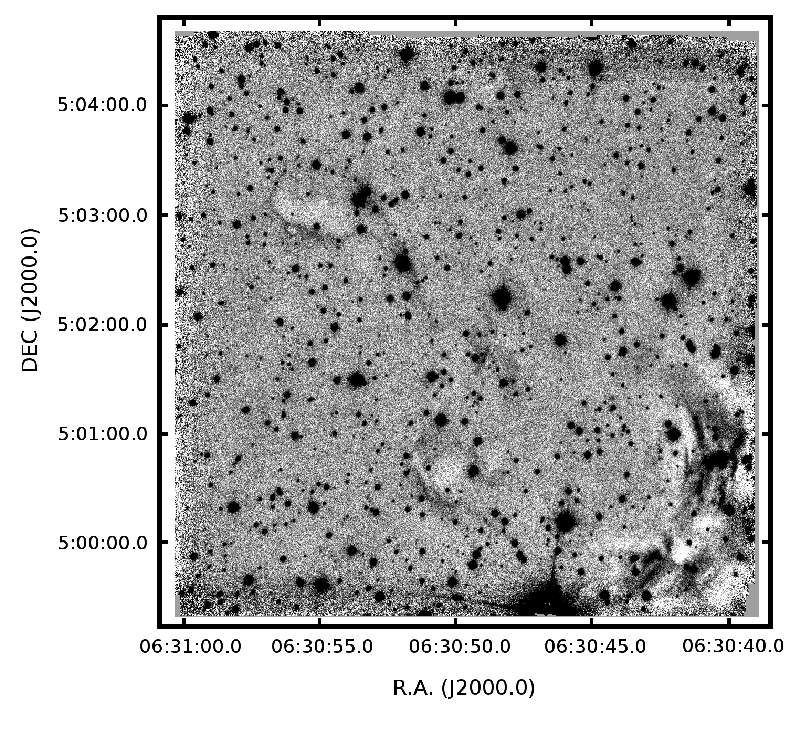}
    \caption{Grayscale image of RN~A in the \Js\ band.}
 \label{fig:s3jsbw}
 \end{figure*}

\begin{figure*}
 \centering
 \includegraphics [width=17cm] {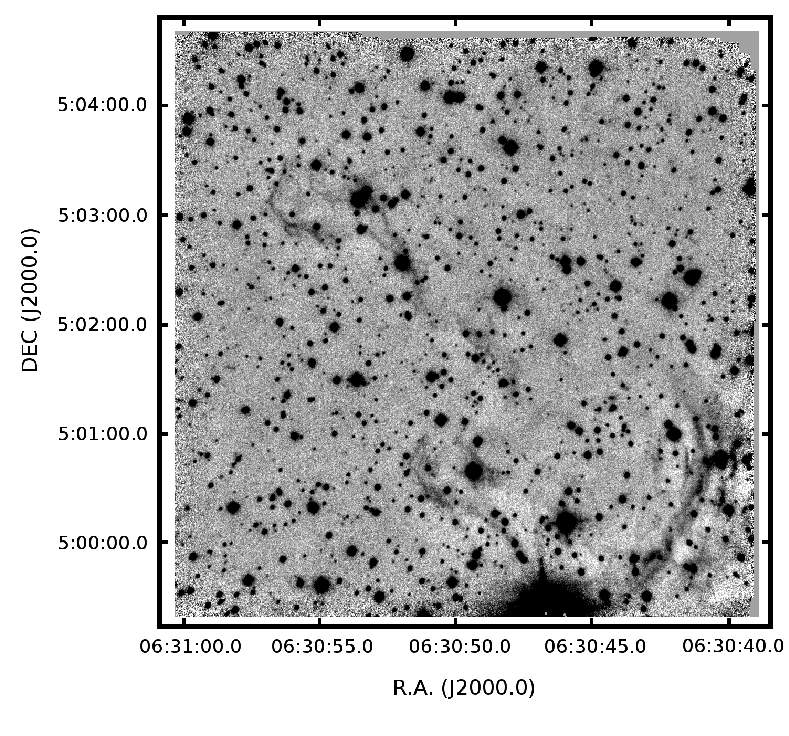}
    \caption{Grayscale image of RN~A in the \H\ band.}
 \label{fig:s3hbw}
 \end{figure*}

\begin{figure*}
 \centering
 \includegraphics [width=17cm] {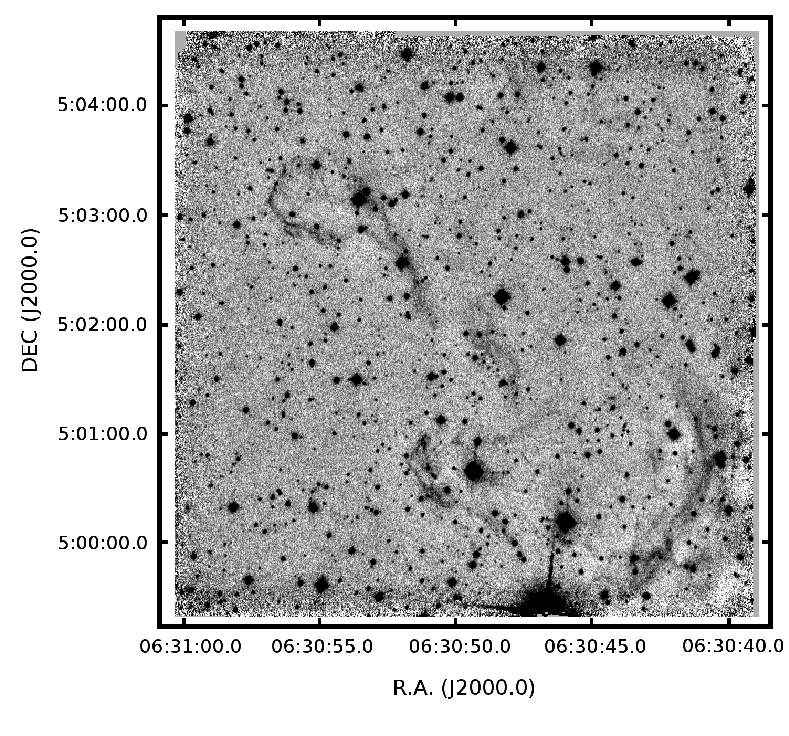}
    \caption{Grayscale image of RN~A in the \Ks\ band.}
 \label{fig:s3ksbw}
 \end{figure*}

 \begin{figure*}
 \centering
 \includegraphics [width=17cm] {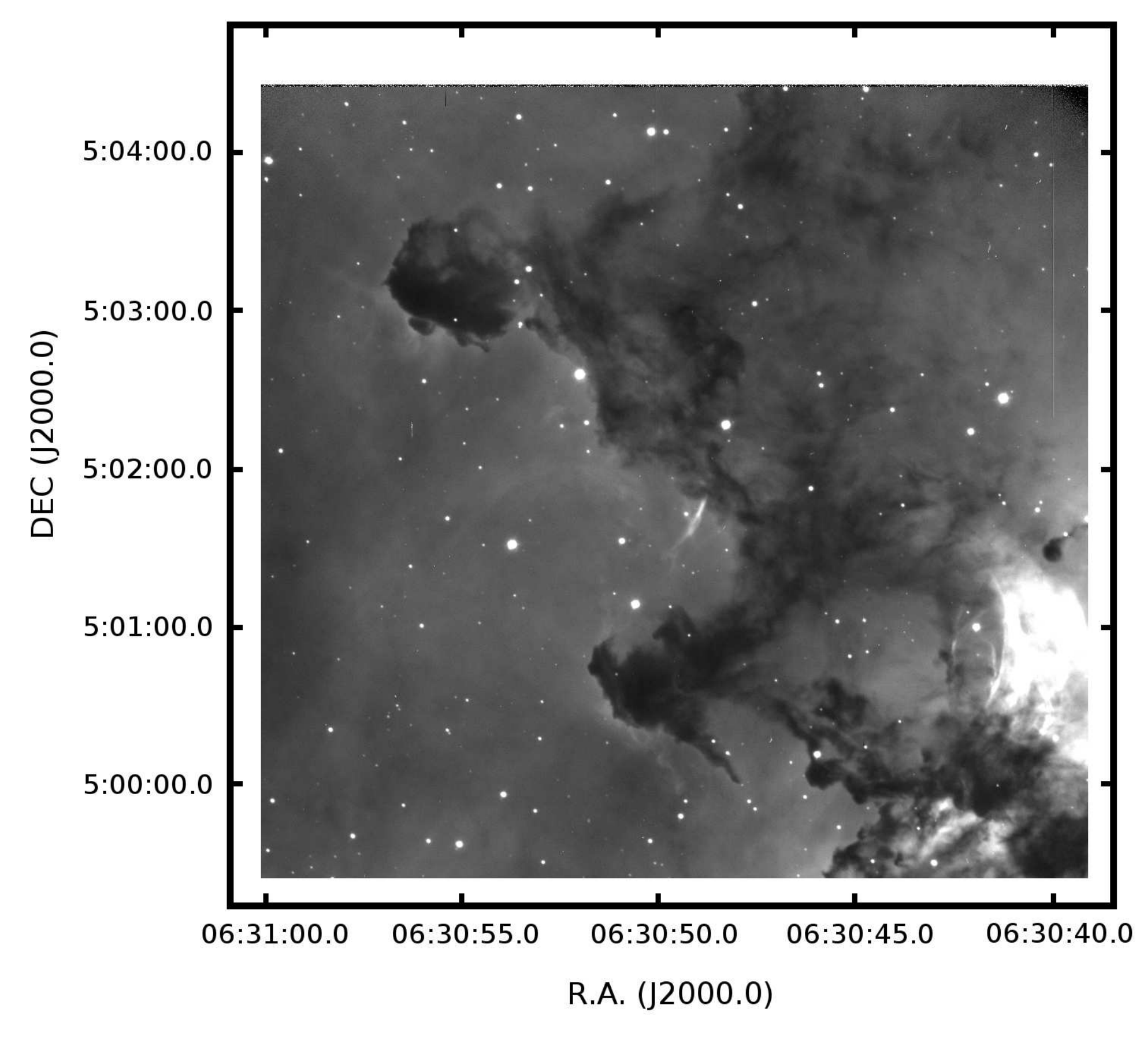}
    \caption{Grayscale image of RN~A in the H$\alpha$ band.}
 \label{fig:s3halpha}
 \end{figure*}

\begin{figure*}
 \centering
 \includegraphics [width=17cm] {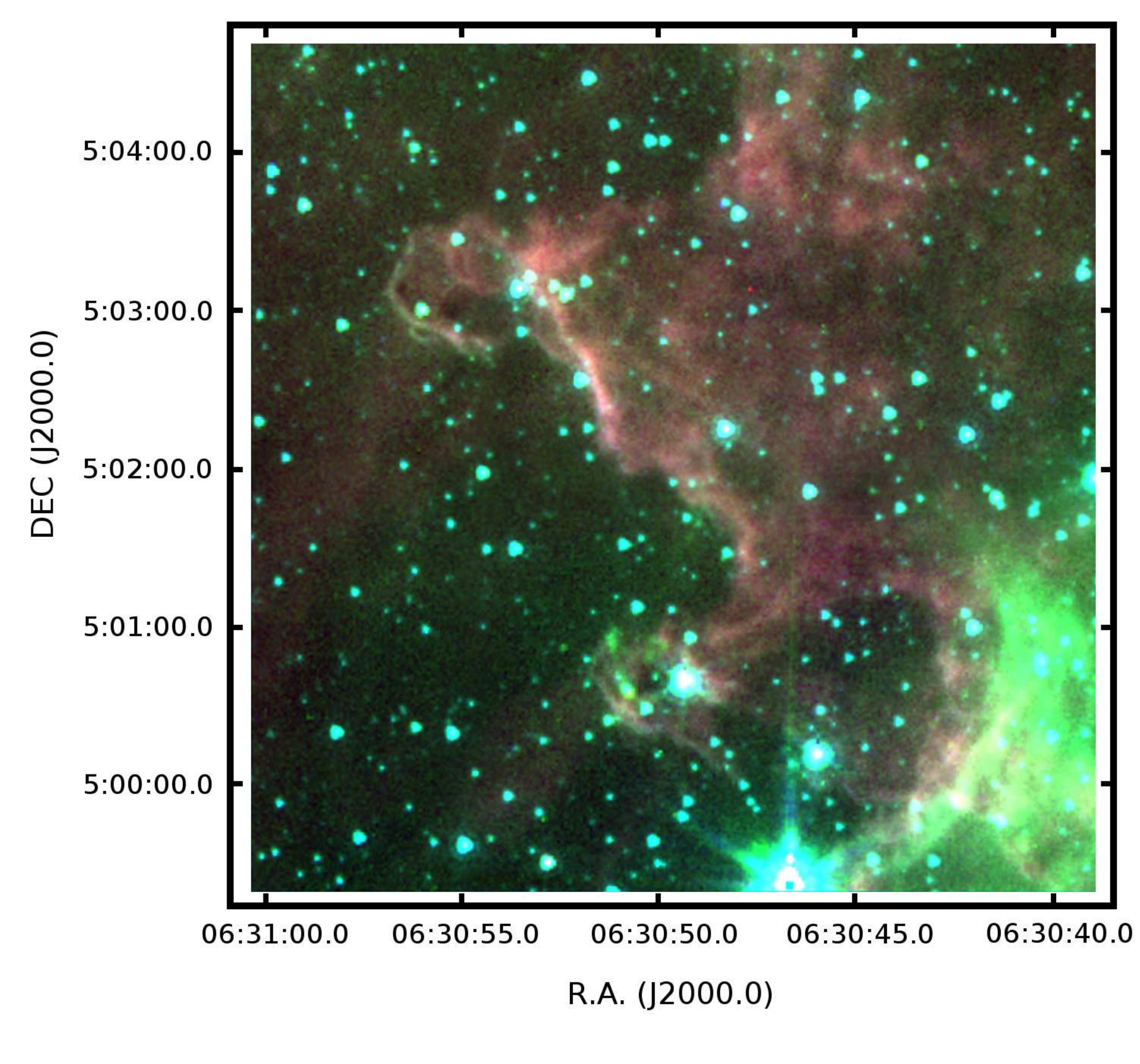}
    \caption{False-color image of the SOFI field in the \textit{Spitzer} IRAC bands. The 3.6, 4.5, and 8.0~$\mum$ bands are coded in blue, green, and red, respectively.}
 \label{fig:s3iracrgb}
\end{figure*}

\begin{figure*}
 \centering
 \includegraphics [width=17cm] {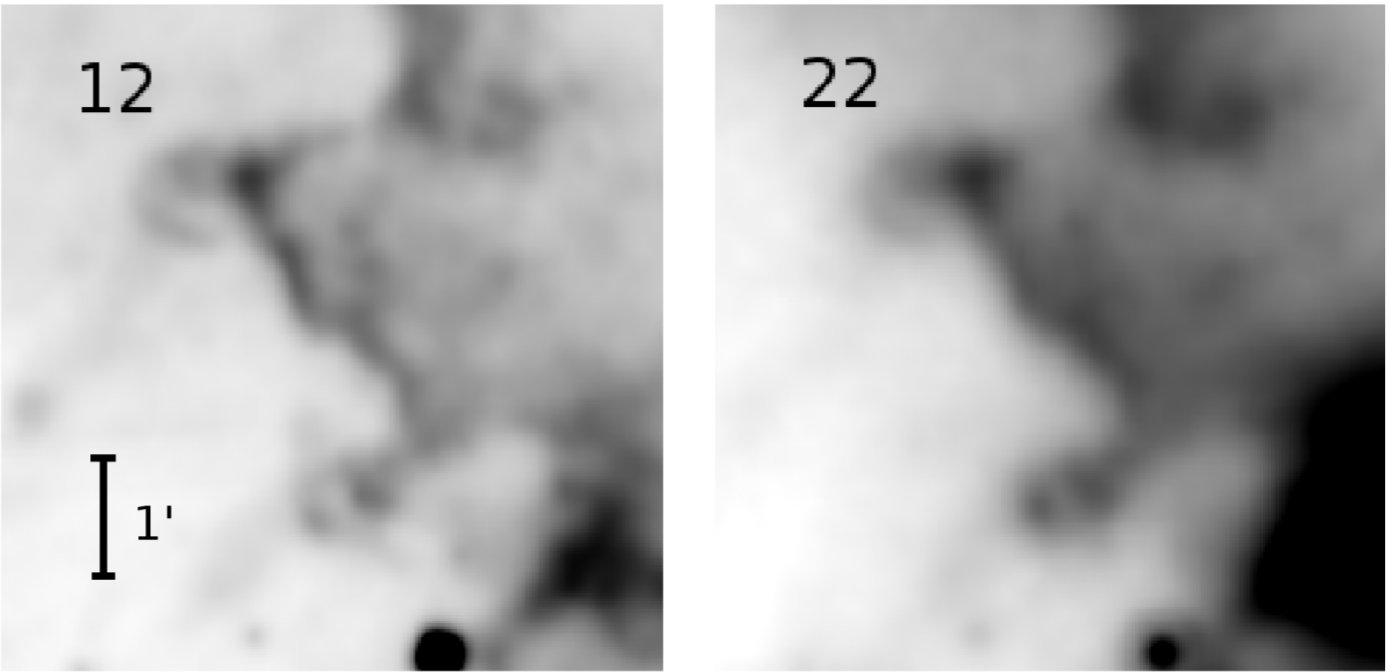}
    \caption{WISE images of the SOFI field (4\farcm9 $\times$ 4\farcm9) in negative grayscale. The numbers in the top lefthand corners indicates the wavelength in microns.}
 \label{fig:s3_wise}
 \end{figure*}

\begin{figure*}
 \centering
 \includegraphics [width=17cm] {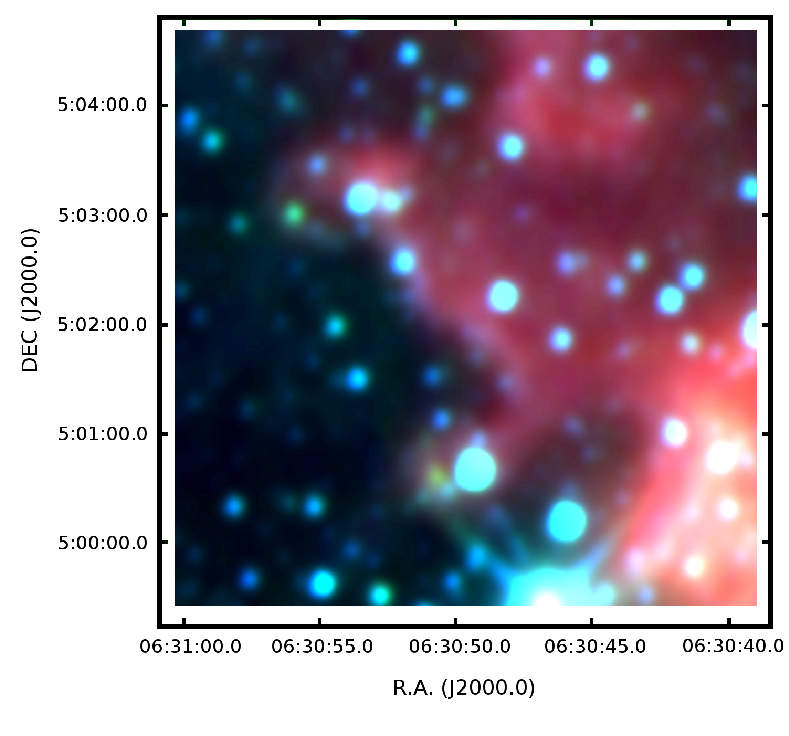}
    \caption{False-color image of the SOFI field in the WISE bands. The 3.4, 4.6, and 22~$\mum$ bands are coded in blue, green, and red, respectively.}
 \label{fig:s3wisergb}
 \end{figure*}

\begin{figure*}
 \centering
 \includegraphics [width=8cm] {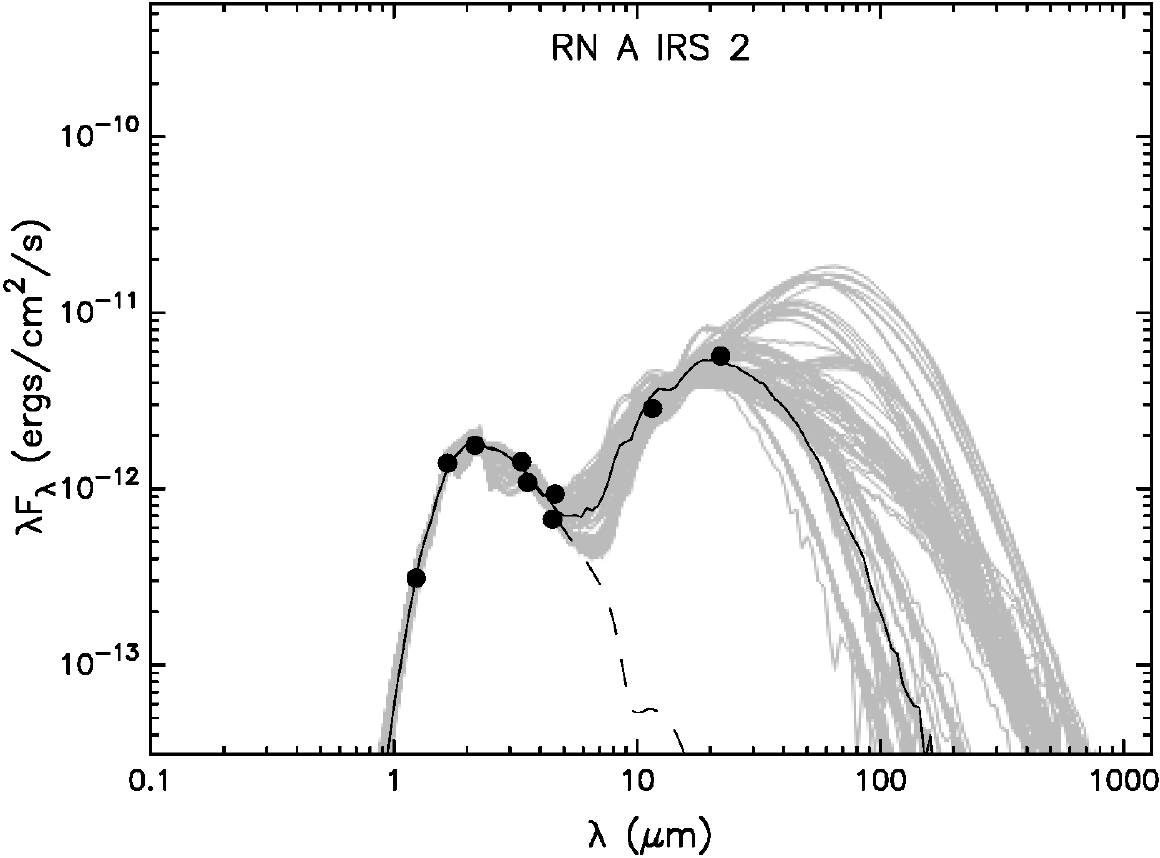}
   \caption{SED for RN~A IRS~2. The circles show the flux values for \jhks, IRAC, and WISE. The best fit is shown with a solid black line, and fits with $\chi_{i}^{2}-\chi_{best}^{2} < 3$ are shown with gray lines. The dashed line is  the stellar photosphere of the best fit including the reddening by interstellar extinction.}
\label{fig:sedfit_acis86}
\end{figure*}

\begin{figure*}
 \centering
 \includegraphics [width=8cm] {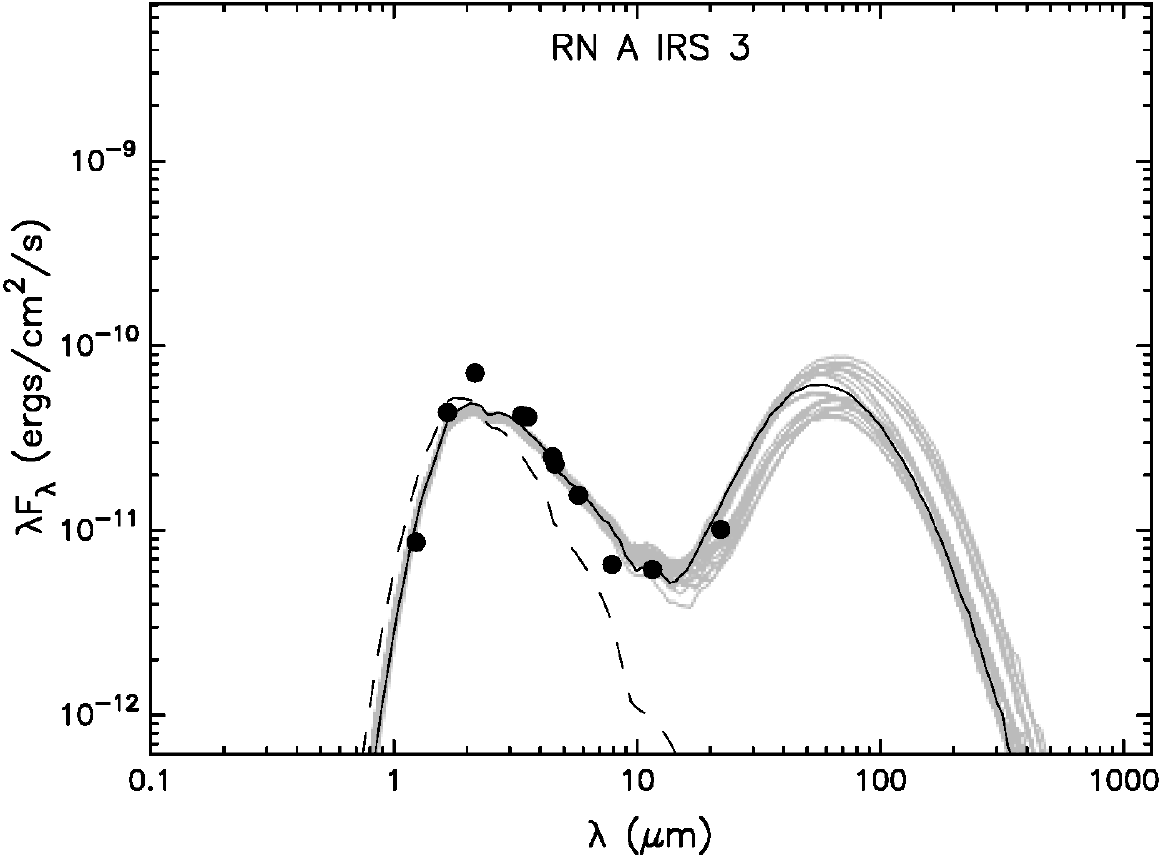}
   \caption{As Fig. \ref{fig:sedfit_acis86}, but for RN~A IRS~3.}
\label{fig:sedfit_rnabrightest}
\end{figure*}

\begin{figure*}
 \centering
 \includegraphics [width=8cm] {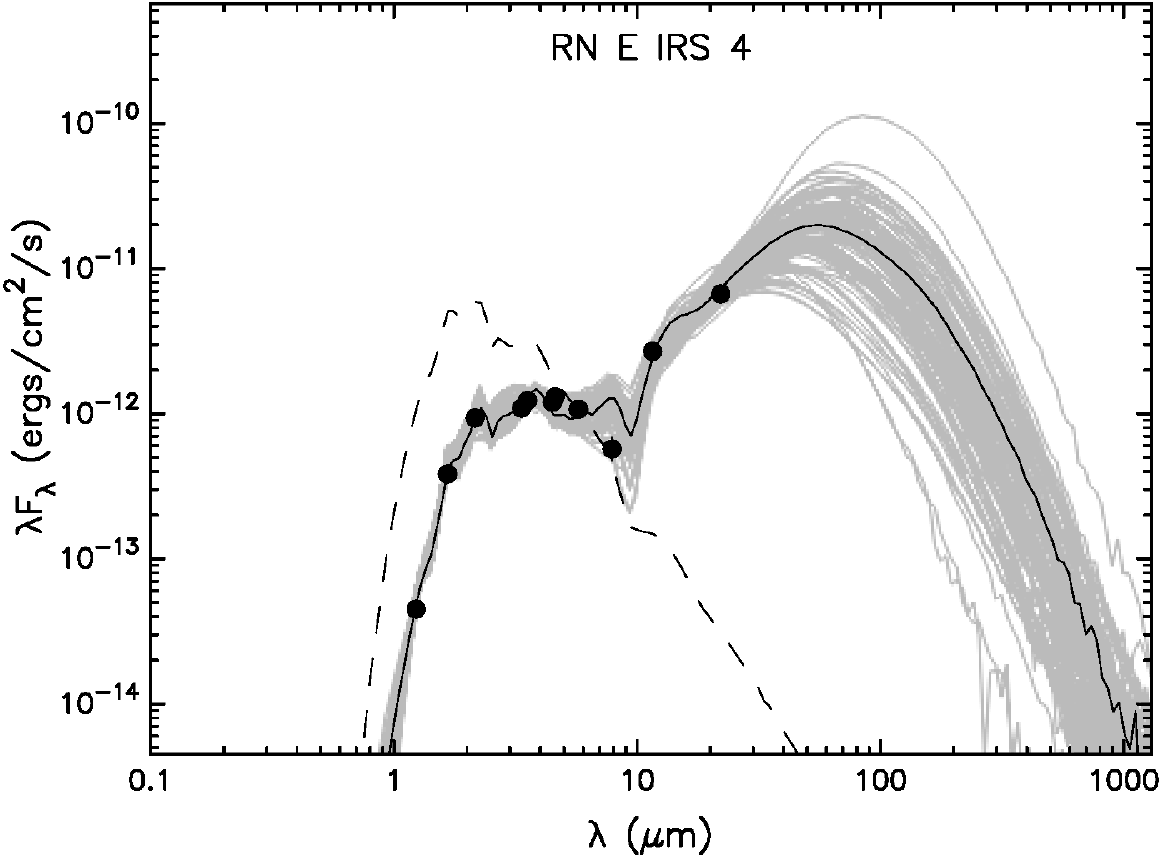}
   \caption{As Fig. \ref{fig:sedfit_acis86}, but for RN~E IRS~4}.
\label{fig:sedfit_npillar}
\end{figure*}

\end{appendix}

\end{document}